\documentclass[aps,pra,onecolumn,superscriptaddress,floatfix]{revtex4-2}

\usepackage[utf8]{inputenc}
\usepackage{amsmath,amsfonts,amssymb}
\usepackage{graphicx}

\usepackage{hyperref}
\usepackage[usenames,dvipsnames]{xcolor}

\hypersetup{colorlinks=true, linkcolor=BrickRed, urlcolor=blue!50!black, citecolor=blue!50!black, pdfproducer={}}

\newcommand{\be}{\begin{eqnarray}}
\newcommand{\ee}{\end{eqnarray}}

\newcommand{\I}{{\bf I}}

\begin{document}
\author{Danial Ghamari}
\affiliation{Physics Department, Trento University, Via Sommarive 14, 38123 Povo (Trento), Italy.}
\affiliation{INFN-TIFPA, Via Sommarive 14, 38123 Povo (Trento), Italy.}
\author{Roberto Covino}
\affiliation{Frankfurt Institute for Advanced Studies, Ruth-Moufang-Straße 1, Frankfurt am Main, D-60438, Germany}
\author{Pietro Faccioli}
\affiliation{Bicocca Quantum Technology Center and Physics Department, University of Milan Bicocca, Piazza della Scienza 2/A, Milan, Italy.}
\affiliation{INFN-TIFPA, Via Sommarive 14, 38123 Povo (Trento), Italy.}
\email{pietro.faccioli@unimib.it}

\title
{  Sampling a rare protein transition with a hybrid classical-quantum computing algorithm.}
\date{\today}


\begin{abstract}
  Simulating spontaneous structural rearrangements in macromolecules with classical Molecular Dynamics (MD) is an outstanding challenge. Conventional supercomputers can access time intervals up to tens of $\mu$s, while many key events occur on exponentially longer time scales. Transition path sampling techniques have the advantage of focusing the computational power on barrier-crossing trajectories, but generating uncorrelated transition paths that explore diverse conformational regions remains an unsolved problem. We employ a path-sampling paradigm combining machine learning (ML) with quantum computing (QC) to address this issue. We use ML on a classical computer to perform a preliminary uncharted exploration of the conformational space. The data set generated in this exploration is then post-processed to obtain a network representation of the reactive kinetics. 
  Quantum annealing machines can exploit quantum superposition to encode all the transition pathways in this network in the initial quantum state and ensure the generation of completely uncorrelated transition paths.
In particular, we resort to the DWAVE quantum computer to perform an all-atom simulation of a protein conformational transition that occurs on the ms timescale. Our results match those of a special purpose supercomputer designed to perform MD simulations. These results highlight the role of biomolecular simulation as a ground for applying, testing, and advancing quantum technologies.
\end{abstract}
\maketitle

\section{Introduction}
Molecular dynamics (MD) simulations yield microscopic insight into macro-molecular systems' structural and dynamical properties. These simulations can, in principle, elucidate the physico-chemical mechanisms involved in complex thermally activated conformational transitions and predict their kinetic and thermodynamic properties. In practice, they are often limited by the path sampling problem, which denotes the challenge of harvesting a statistically significant number of trajectories that connect given initial and final conformational states by overcoming considerable free-energy barriers ~\cite{revES}. MD simulations invest an exponentially large fraction of computational time in simulating thermal fluctuation in metastable states, and struggle to characterize these barrier-crossing events.
To tackle this challenge, various techniques have been developed over the past few decades~\cite{revES}.

Among them, Transition Path Sampling (TPS)~\cite{TPS1,TPS2} has the advantage of focusing directly on generating productive trajectories, given only the definition of two conformational states. It does not introduce unphysical forces to accelerate the sampling or drive the system, nor involves any collective variable choice. 

TPS is a Monte Carlo scheme in which trial trajectories are generated using MD simulations initiated in the transition region and then accepted or rejected according to a Metropolis criterion. The acceptance rate strongly depends on the choice of initial conditions from which the MD simulations are initiated. Low acceptance rates often limit the simulation of complex transitions. To tame this issue, a promising reinforcement learning scheme has been introduced to optimize this procedure \cite{CovinoAI}, enabling the simulation of rare and complex processes, such as membrane-protein assembly. 
However, TPS still requires sampling the duration of the transition with unbiased simulations, which can be very expensive. 
For example, generating a \emph{single} trial trajectory in a protein folding simulation would require running $\mu$s long MD simulations.

\begin{figure}[t!]
\centering
\includegraphics[width=.9\textwidth]{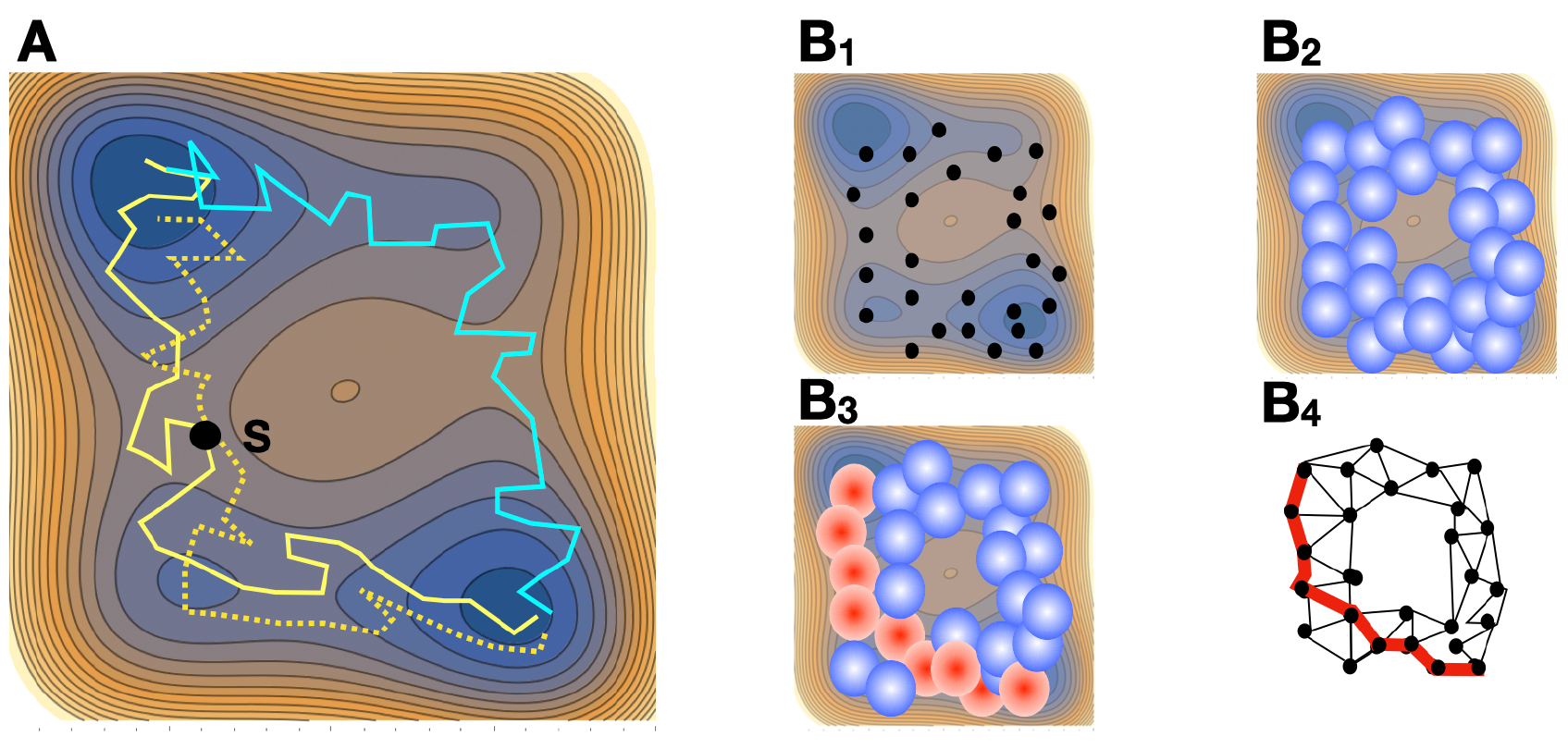}
\caption{ Panel A: Schematic representation of a shooting move in the conventional TPS algorithm. Panels B$_1$-B$_4$: illustration of our path sampling scheme that combines ML and MD performed on a classical computer with QC performed on a quantum annealing machine: First, a discrete ensemble of configurations is generated using the iMapD uncharted manifold exploration algorithm (B$_1$). Each point is assigned to a hyperspherical region of configuration space. The radius of the hyperspheres is chosen comparable with the average inter-configuration distance (B$_2$). Next, a formalism combining stochastic path integrals with renormalization group theory is used to obtain a graph representation of the kinetics (B$_3$). In this theory, the probability reaching the product by visiting a given sequence of finite regions can be calculated from the sum of edges of the an undirected network (B$_4$). }
\label{fig_gtps}
\end{figure}

A second problem that can potentially limit the applicability of TPS is the presence of long autocorrelation in its Markov chain, illustrated by the example shown in Fig.~\ref{fig_gtps}. In this simple two-dimensional energy landscape, the transition region between two states can occur through two distinct transition channels. 
In one of the most efficient formulations of TPS, a new trial path is generated by randomly picking a configuration $S$ in the last accepted path and then computing two independent stochastic trajectories, starting from $S$. 
The trajectory obtained by joining these two trajectories is a viable transition path if it directly connects the reactant and product region. 
Suppose now that this Markov chain is initiated from a trajectory that travels along the leftmost channel in panel A of Fig.~\ref{fig_gtps}. If the energy barrier separating the two channels is large, exploring the rightmost transition channel would require many Monte Carlo steps. 

 We recently proposed to overcome this challenge by integrating molecular simulations, machine learning (ML), and quantum computing (QC) \cite{QTPS}. In such a scheme --hereby referred to as "graph Transition Path Sampling" (gTPS)--, MD simulations driven by ML are employed on a classical computer to efficiently perform a preliminary uncharted exploration of the relevant region of the molecule's configuration space (the so-called intrinsic manifold) (see panel B$_1$ of Fig.~\ref{fig_gtps}).

Each configuration generated during this exploration identifies a finite region of configuration space of size $\sigma$ comparable with the average nearest-neighbor configuration distance (panel B$_2$ of Fig.~\ref{fig_gtps}). 
A rigorous mathematical expression derived by assuming an underlying Langevin dynamics yields the probability for the system to visit a given ordered sequence of these finite regions, in a transition from a given initial and final region (the red path in panel B$_3$ of Fig.~\ref{fig_gtps}). This microscopically reversible coarse-grained theory can be represented by an \emph{undirected} graph, in which the nodes identify the different finite regions, while the probability of a given reactive path is computed from the sum of the edges that connect the initial and final nodes on the network ( panel B$_4$ of Fig.~\ref{fig_gtps}).

In principle, transition paths in a graph may be sampled with classical computers with a discrete version of the conventional TPS, where the trial trajectories are stochastically generated, e.g., by kinetic Monte Carlo. However, for sufficiently large graphs, the resulting Markov chains would be liable to the same autocorrelation problem that appears in the original TPS. This issue can be overcome by resorting to a quantum annealing machine ~\cite{DAS_CHAKRABARTI,Das2008,Albash2018,VenegasAndraca2018,Hauke2020} to generate the trial paths in the Markov chain. In such an approach, transition paths can be encoded in a quantum computer by assigning qubits to each node and edge in the graph (panel A in Fig.~\ref{fig_qc_encode}).
The main point in using a QC approach is that the initial state of the quantum computer can be easily prepared in order to simultaneously encode all the transition paths connecting two given points on the graph (panel B in Fig.~\ref{fig_qc_encode}). Then, a process called adiabatic switching can be applied to promote the components of the computer's wave function associated with the most statistically significant transition paths. 
This way, the quantum annealing machine generates a new uncorrelated trial path at each measurement, i.e., a productive transition pathway with high statistical weight. However, in general, in realistic conditions (e.g., for suboptimal choices of the time employed to perform the adiabatic switching procedure), the resulting path distribution may not exactly correspond to that of the underlying Langevin dynamics. However, a straightforward Metropolis acceptance/rejection criterion can be implemented on a classical computer to correct for such a deviation.

Any time the quantum computer is reset in its initial state, the system loses memory of the previously generated path, so the trajectory obtained at the end of the annealing process is completely uncorrelated to the previous one in the Markov chain. 

While resorting to quantum superposition overcomes the autocorrelation problem, the uncharted exploration implemented on the classical computer exploits a data-driven scheme \cite{IMAPD} to enhance the probability of diffusing into previously unexplored regions, thus promoting the chance of discovering multiple transition channels.
The key question is whether the existing quantum computers can accurately and realistically simulate transitions too complex to be investigated by plain MD, even on large GPU computer clusters. 

In this work, we demonstrate that the answer is affirmative: we use only a few GPUs on a classical computer cluster and a few hundred qubits on the DWAVE quantum computer to characterize, with atomic resolution, a rare conformational transition that occurs on a time scale of a few ms in the native state of Bovine basic Pancreatic Trypsin Inhibitor (BPTI) protein (PDB code: 5PTI). While this
dynamics is unattainable by plain MD simulations on conventional computer clusters, it has been fully characterized using the Anton special purpose supercomputer \cite{Anton2010}, the result of which can be used to benchmark our approach.
gTPS enables us to generate configurations that can be compared with those found in the metastable states discovered by Anton's simulations. In addition, gTPS yields several fully uncorrelated transition paths connecting two configurations belonging to different states, which can be compared with the free-energy landscape obtained by plain MD. 

\section{Theory}

The implementation of the gTPS scheme within our hybrid classical-quantum computing paradigm involves three key steps: (i) the preliminary uncharted exploration of the relevant macromolecular configurations---the intrinsic manifold---, (ii) the construction of a coarse-grained network representation of the stochastic dynamics from a data set harvested from the results of molecular simulations, and (iii) the quantum encoding of the transition path sampling problem on a QC. 

In the following, we review these three steps, focusing on the conceptual aspects and referring the reader to the original publication and Supplementary Information (\textrm{SI}) for all mathematical derivations.

\subsection*{Step 1: Uncharted exploration of the relevant configurations}\label{uncharted}
 
The Intrinsic Map Dynamics (iMapD) algorithm~\cite{IMAPD}, schematically illustrated in panel A of Fig.~\ref{fig_polar}, provides an efficient scheme to perform an uncharted exploration of the intrinsic manifold: Starting from some local sampling performed by plain MD, iMapD relies on diffusion maps \cite{DMAP} to obtain a low dimensional representation of the molecular configurations $X_i$ and identify the boundary of the explored portion of the intrinsic manifold. To generate new configurations outside the boundary of the explored region, we first identify 
the direction perpendicular to the boundary surface at each boundary point $X_i^b$. 
This is done by computing an average over the coordinates of the configurations in the neighborhood of $X_i^b$ (leading to $X_{\textrm avg}^i$) and then performing a translation in the direction informed by the difference $X_i^b- X_{avg}^i$. This translation is applied after projecting the configurations onto the dominant subspace in a Principal Component Analysis (PCA), in order to account for the local structure of the intrinsic manifold (see section S1 in the \textrm{SI} for further detail). 

The new configurations outside the boundary of the explored region are used as initial conditions to start a new round of local MD sampling. By iterating this procedure, one continuously finds new configurations in previously unexplored regions of the intrinsic manifold. 

\begin{figure}[t!]
\centering
\includegraphics[width=.9\textwidth]{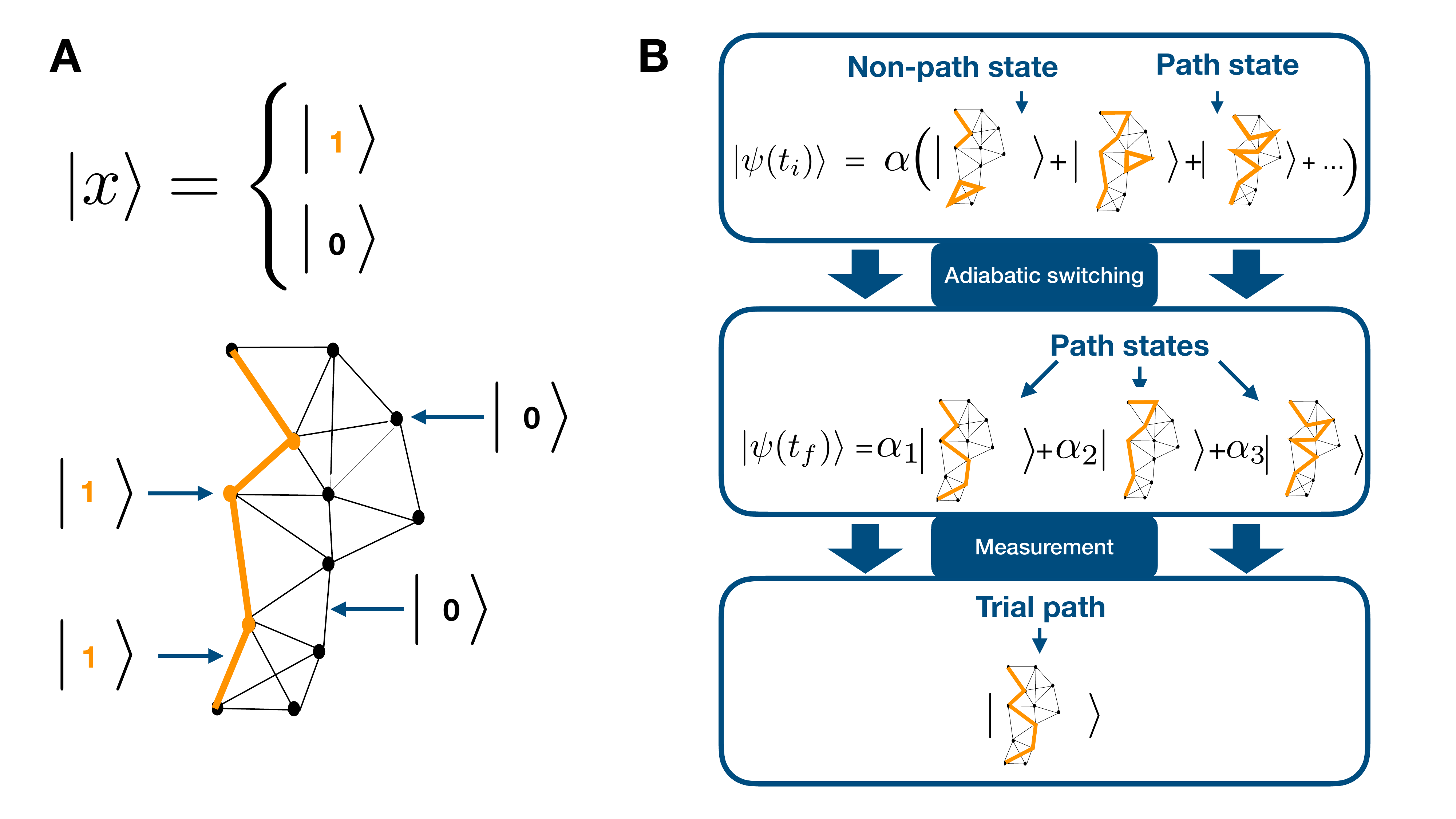}
\caption{Left panel: A schematic illustration of a quantum encoding of the path sampling problem based on Quadratic Unconstrained Binary Optimization. Right panel: illustration of how a (quasi)- adiabatic switching procedure on a quantum annealing machine yields uncorrelated trial paths on the graph.}\label{fig_qc_encode}
\end{figure}

\subsection*{Step 2: Coarse-grained representation of the stochastic dynamics on the intrinsic manifold}\label{coarsegraining}

From the large set of molecular frames generated during the uncharted exploration, we harvest a representative set $\mathcal{C}=\{X_k\}_{k=1,\ldots,\nu}$. These configurations are selected to be distributed as uniformly as possible on the intrinsic manifold (see section S2 of the \textrm{SI}), thus are not sampled from a Boltzmann distribution. However, they can still be employed to build a coarse-grained representation of the stochastic dynamics that obeys the correct detailed balance condition.

The key idea is to regard each configuration $X_k\in \mathcal{C}$ as the representative element of a finite region of configuration space $R_k$ of size $\sigma$. By choosing $\sigma$ comparable with half the average distance between nearest-neighbor configurations in $\mathcal{C}$, the union of all finite regions $R= \bigcup_{k=1}^\nu R_k$ covers the explored portion of the intrinsic manifold (as in panel B$_{2}$ of Fig.~\ref{fig_gtps}). The size of the finite regions $\sigma$ defines the maximum spatial resolution scale at which the dynamics can be represented, given the structural information encoded in the finite data-set~$\mathcal{C}$.

In Ref.\cite{QTPS}, we derived a rigorous mathematical formalism to build from $\mathcal{C}$ an effective theory for the stochastic dynamics on the intrinsic manifold, assuming an underlying microscopic Langevin dynamics. In this theory (whose main equations are collected in section S3 of the \textrm{SI}), the short-distance dynamics of the microscopic theory is smeared out to match the resolution scale $\sigma$. This is done by pre-computing local averages in configuration space, using $\nu$ Gaussian distributions of variance $\sigma$, centered at the configurations $\{X_k\}_{k=1,\ldots,\nu}$ (for details about this derivation, we refer the reader to the original publication.) This effective theory's maximum temporal resolution scale $\Delta t$ is defined by the average time to cover a distance $\sigma$ in configuration space in the underlying microscopic dynamics.

A transition pathway corresponds to given sequence of finite regions $R_{i_0}\to R_{i_2} \to \ldots \to R_{i_N}$ that are crossed in order, while going from the reactant $i_0$ to the product $i_N$ (illustrated by the red path in panel B$_{3}$ of Fig.~\ref{fig_gtps}). Hence, a transition path in the coarse-grained theory is specified by the integer vector $\I = (i_1, \ldots, i_{N_I})$, where $i_k$ is the region visited at time $t_k$. 
The expression for the conditional probability to perform a transition from a given initial region $i_0$ to a given final region $i_N$, in a time interval $t$ in the effective theory reads: 
\begin{eqnarray}\label{Pcg}
P_{cg}(i_N, t| i_0) \propto e^{t s_0} \sum_{i_1,\ldots,i_{N-1}} e^{-S[\I]},
\end{eqnarray}
where $\I=(i_1, \ldots,i_{N-1})$ denotes a path and $s_0$ is a frequency cut-off scale which is set of the order of the average rate of escape from the finite regions.
$S(\I)$ is called the coarse-grained Hamilton-Jacobi action and reads 
\begin{eqnarray}\label{S}
 S(\I) &=& \sum_{k=0}^N w_{i_{k+1} i_k}, \qquad \textrm{where}\\
 \label{wijt}
w_{i j} &=&\frac{|X_i-X_j|}{2 \sqrt{D_{cg}}} (L_i +L_j), \quad L_i = \sqrt{\frac{ \text{V}_\textrm{eff}^\textrm{cg}(i)}{2m D_{cg}}+s_0}.
\label{wijs0}
\end{eqnarray}
In these equations, $m$ is an average atomic mass, $D_{cg} = \frac{s_0 \sigma^2}{2}$ is called the coarse-grained effective diffusion constant and
$ \text{V}_\textrm{eff}^\textrm{cg}(i)$ is the so-called coarse-grained effective potential and contains the information about the local diffusive kinetics. Indeed, $ \text{V}_\textrm{eff}^\textrm{cg}(i)$ is directly related to the probability for the system to remain in the same finite region in an infinitesimal time interval $\Delta t$, namely 
\begin{eqnarray}
 P_{cg}(i, \Delta t| i) \propto e^{- \frac{\Delta t}{2 m D_{cg}} \text{V}_\textrm{eff}^\textrm{cg}(i)}.
\end{eqnarray} 
Hence, this equation provides a straightforward way of computing the coarse-grained effective potential by generating many short MD trajectories starting from configurations in the region $i$ and counting how many of them remain in the same region after a time interval $\Delta t\sim 1/s_0$. Note that the combination $\frac{2 m D_{cg}}{ \text{V}_\textrm{eff}^\textrm{cg}(i)}$ can be identified with the average "life-time" $\tau_i$ of the region $i$. Hence, $ \text{V}_\textrm{eff}^\textrm{cg}(i)$ can be alternatively computed by measuring the average time to reach a point at a root-mean-square deviation (RMSD) $\sigma$ from the centroid of region $R_i$. Note that this calculation could also be performed on-the-flight, during the iMapD exploration.

The dynamics of our coarse-grained stochastic theory can be represented on an undirected network. To this end, each finite region $R_k$ is assigned to a node, while the edge connecting the nodes $i$ and $j$ is assigned a weight $w_{ij}$ given by Eq.~(\ref{wijs0}). This way, the probability of a given coarse-grained path $\I$ is evaluated by the negative exponent of the sum of the weights of all the edges forming the path $I$ (see panel B$_4$ in Fig.~\ref{fig_gtps}). 

Let us now discuss the kinetic information that can be extracted from the paths calculated using the gTPS formalism. As shown in the \textrm{SI}, the time to perform a transition from the initial to the final node can be estimated as $t = N \langle \tau \rangle$, 
 where $\langle \tau \rangle= \frac{1}{\nu}\sum_{i=1}^\nu \tau_i$ is the average node life-time and $N$ is the number of steps taken by the path.

In an approximation in which all links $w_{ij}$ are set equal to their average value $\bar{w}$, the probability of a given path crossing $N$ nodes is $P\sim e^{- \bar{w} N}$. Hence, in this mean-field approximation, the most probable trajectory corresponds to the shortest path between the initial and final node and the corresponding duration $t$ provides a lower-bound estimate of the shortest time at which this transition can occur in equilibrium conditions. It is important to emphasize that the gTPS formalism does not capture the exponential time spent by the system to explore metastable regions, thus can only be used to estimate the transition path time for conformational changes that involve crossing a single free-energy barrier.

This feature represents an important difference between the gTPS theory and the Markov State Model (MSM) approach since the latter yields a complete representation of the relaxation kinetics. On the other hand, defining a MSM is significantly more computationally expensive, since it involves computing the full transition matrix. Instead, in gTPS, it is sufficient to estimate the life-times of the $\nu$ nodes in the graph. 

\subsection*{Step 3: Quantum computing of transition paths}

Our quantum encoding of the transition path sampling problem is based on assigning quantum bits (qubits) to each node and each edge of the graph (Fig.~\ref{fig_qc_encode} A). Since qubits are two-level quantum systems, their state can be conventionally labeled using the eigenstates of the spin 1/2 operator, $\hat \sigma_z$. In the following, we denote with $|s^i\rangle$ the qubit associated with the $i-$th node in the graph and with $|s^{ij}\rangle$ the qubit associated to the edge connecting the nodes $i$ and $j$. 

Let us now review how it is possible to use a quantum annealing machine to obtain qubits configurations corresponding to transition paths between given initial and final nodes. To define this procedure, we introduce two quantum Hamiltonians (explicitly reported in the \textrm{SI}) that act on the Hilbert space of our set of qubits: $\hat H_\textrm{i}$ and $\hat H_\textrm{f}$. $\hat H_\textrm{i}$ is used to prepare the initial state of the quantum computer. In particular, we are interested in initializing each qubit in an equal linear superposition $\frac{1}{\sqrt{2}}\left(|s=1/2\rangle+|s=-1/2\rangle\right)$. This condition is fulfilled by choosing an initial Hamiltonian that is defined in terms of Pauli $\hat \sigma_x$ operators \cite{DAS_CHAKRABARTI,Das2008,Albash2018,VenegasAndraca2018,Hauke2020}.

The final Hamiltonian $H_\textrm{f}$ is defined in such a way to contain a gap in the spectrum, with the low-energy sector filled by so-called "path states," i.e., qubit configurations that represent transition paths. All other 
eigenstates above the gap are associated with graph configurations that do not represent transition paths (see Fig.~\ref{fig_qc_encode}). Furthermore, $\hat H_\textrm{f}$ must be chosen in such a way that the energy of the state corresponding to a given transition path $\I$ coincides with the path action $S(\I)$ defined in Eq.~(\ref{S}). 
With this choice, the low-lying states of $\hat H_\textrm{f}$ are ordered in terms of their relative probability in the effective stochastic action defined in step 2. Any algorithm capable of intercepting low-energy states will yield highly probable transition paths. 

After being prepared in the ground-state of $\hat H_i$, the system is subjected to a time-dependent Hamiltonian 
\be\hat H(t) = A(t)~\hat H_{\textrm{in}} + B(t)~\hat H_f,\ee
with scheduling functions $A(t)$ and $B(t)$. These are chosen such that initially $A(0) = 1$ and $B(0) = 0$, while at the end of the protocol, i.e., at $t=t_{\textrm{sweep}}$, one has $A(t_{\textrm{sweep}}) = 0$ and $B(t_{\textrm{sweep}}) = 1$. That is, the sweep starts with $H(0)=H_\textrm{i}$ and ends in $H(t_{\textrm{sweep}})= H_\textrm{f}$. 

In an ideal closed system, the adiabatic theorem implies that if the sweep is performed sufficiently slowly as compared to the minimal energy gap $\Delta E$ between the instantaneous ground state and the first excited state of $H(t)$ ( i.e., for $t_{\textrm{sweep}}\gg \hbar/\Delta E$) then the system remains in its ground-state, thus reaching the ground-state of $\hat H_\textrm{f}$ at the end of the sweeping process.
Therefore, in ideal conditions, this adiabatic annealing process would systematically yield the least-action path $\bar{\I}$~\cite{QDRP}, also called the dominant transition pathway \cite{DRP1}.

In practice, quantum annealing machines are coupled to their environment and the sweeping time is not arbitrarily slow. As a result, any sweeping process lasting a finite time $t_\textrm{sweep}$ has a finite probability $P(\I|t_\textrm{sweep})$ to yield a different path $\I\ne \bar \I$ . 
\begin{figure}[t!]
\centering
\includegraphics[width=.9\textwidth]{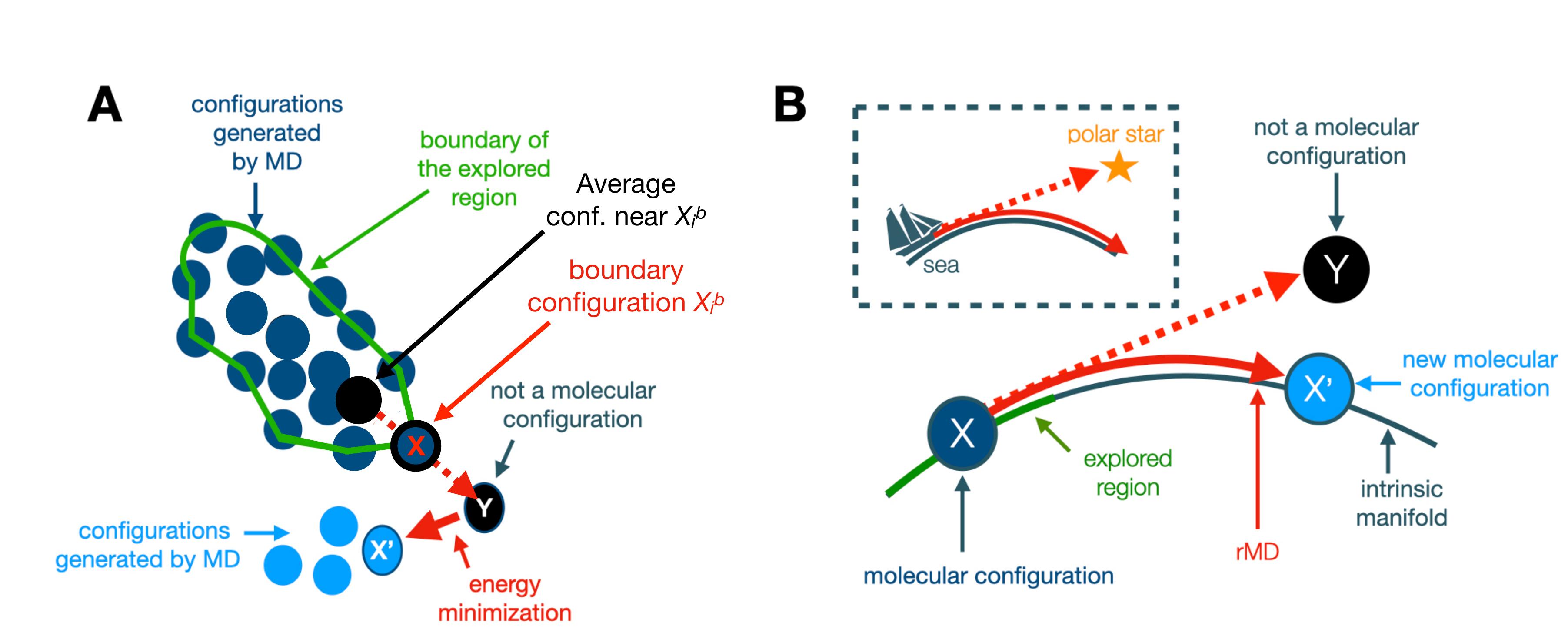}
\caption{(A) Schematic illustration of the iMapD algorithm. (B) Schematic illustration of the Polar Star shooting move. The name arises from the analogy of sailors' navigating system based on pointing to far distant stars to guide the navigation on the Earth's surface (figure in the inset). }\label{fig_polar}
\end{figure} 

For sufficiently long sweeping times, all paths generated this way will have low action $S(\I)$, thus will correspond to statistically significant reactive trajectories in the underlying coarse-grained theory. 
A reweighting Metropolis scheme can be additionally imposed to enforce a detailed balance condition, which ensures that the ensemble of paths is distributed according to the $e^{-S(\I)}$ probability \cite{QTPS}.

\section{Results}\label{Results}

\subsection{Improved uncharted exploration}

Using iMapD to navigate the energy landscape of biological macromolecules can be computationally challenging. In the following, we introduced an improvement in this algorithm, which enabled us to rapidly explore the region of the intrinsic manifold of protein BPTI close to its native state. 

The iMapD algorithm's efficiency depends on the computational cost of generating new viable molecular configurations outside the explored region. To this end, iMapD uses a procedure we shall refer to as "Euclidean scheme", which consists in picking a configuration $X$ near the center of the visited region, measuring its distance $d$ from the border of the explored region, and adding to $X$ a multi-dimensional vector to translate it out the explored region, using principal component analysis to keep into account of the local structure of the configurational dataset near $X$ (see panel A in Fig.~\ref{fig_polar}). 
A problem with this procedure is that the new set of atomic coordinates $Y$ generally does not represent a viable molecular configuration. For example, it violates the topological constraints dictated by the molecule's chemical structure (as illustrated in panel B of Fig.~\ref{fig_polar}). A viable molecular configuration $X'$ may be obtained by running an energy minimization starting from $Y$. However, to ensure the stability and reliability of this procedure, one is forced to take incremental shooting moves. 

To overcome this issue, we introduced an improved procedure (that we refer to as the "Polar Star" scheme, see \textrm{SI}), which is schematically illustrated in panel B of Fig.~\ref{fig_polar}. 
The key idea is to use the set of coordinates $Y$ generated after the Euclidean scheme to compute the entries of a contact map $C^0_{ij}$. Next, a specific type of biased simulation called Ratchet-and-pawl MD \cite{rMD1, rMD2, irMD} (rMD) is employed to drive the MD simulation toward a viable molecular configuration $X'$, outside of the explored region. 
In a rMD simulation, the system evolves as in plain MD as long as it spontaneously progresses towards configurations with a larger overlap between the instantaneous and the target contact map. Conversely, a harmonic biasing force switches on to prevent decreasing this overlap. 
This way, the macromolecule rapidly reaches configuration $X'$ with a contact map as close as possible to $C^0_{ij}(Y)$.

We emphasize that the bias in this rMD is not chosen heuristically by the user. Instead, it is determined by the algorithm in an unsupervised way based on the structure of the data-set generated in the previous iterations. Therefore, the exploration remains fully uncharted.

The Polar Star scheme ensures the generation of viable molecular configurations well outside the explored regions, thus significantly improving the efficiency of the exploration. In addition, we employ a combination of moves with different amplitude (i.e. with different choices of the parameter $c$ defined in Eq.~(S2)). Integrating the results of iMapD simulations at different values of $c$ enables us to cope with the existence of structure at different length scales in the macromolecule's energy landscape.

\subsection{Native conformational dynamics of the BPTI protein}

In its native state, the protein BPTI undergoes thermally activated structural rearrangements on the millisecond scale (ms), which cannot be covered with plain MD simulations on conventional high performance computing facilities. However, these transitions have been observed and characterized by means of ms long MD simulations performed on the Anton special purpose supercomputer \cite{Anton2010}. Two of their longest-lived states have also been found using a Boltzmann generator, an example of generative artificial intelligence applied to sampling macromolecular structures \cite{NoeScience}.

{\bf Exploration of the intrinsic manifold near the native state.}
First, we addressed whether the iMapD algorithm can discover protein conformations close to those observed in the Anton simulations \cite{Anton2010}. 
Using the conventional iMapD scheme based on the Euclidean scheme, we could not reach configurations with an RMSD distance from the initial native structure comparable to that observed in plain MD simulations. Conversely, the Polar Star scheme enabled us to complete this task within a cumulative simulation time of $\sim 3\mu s$. 

\begin{figure}[t!]
\includegraphics[width=0.9\textwidth]{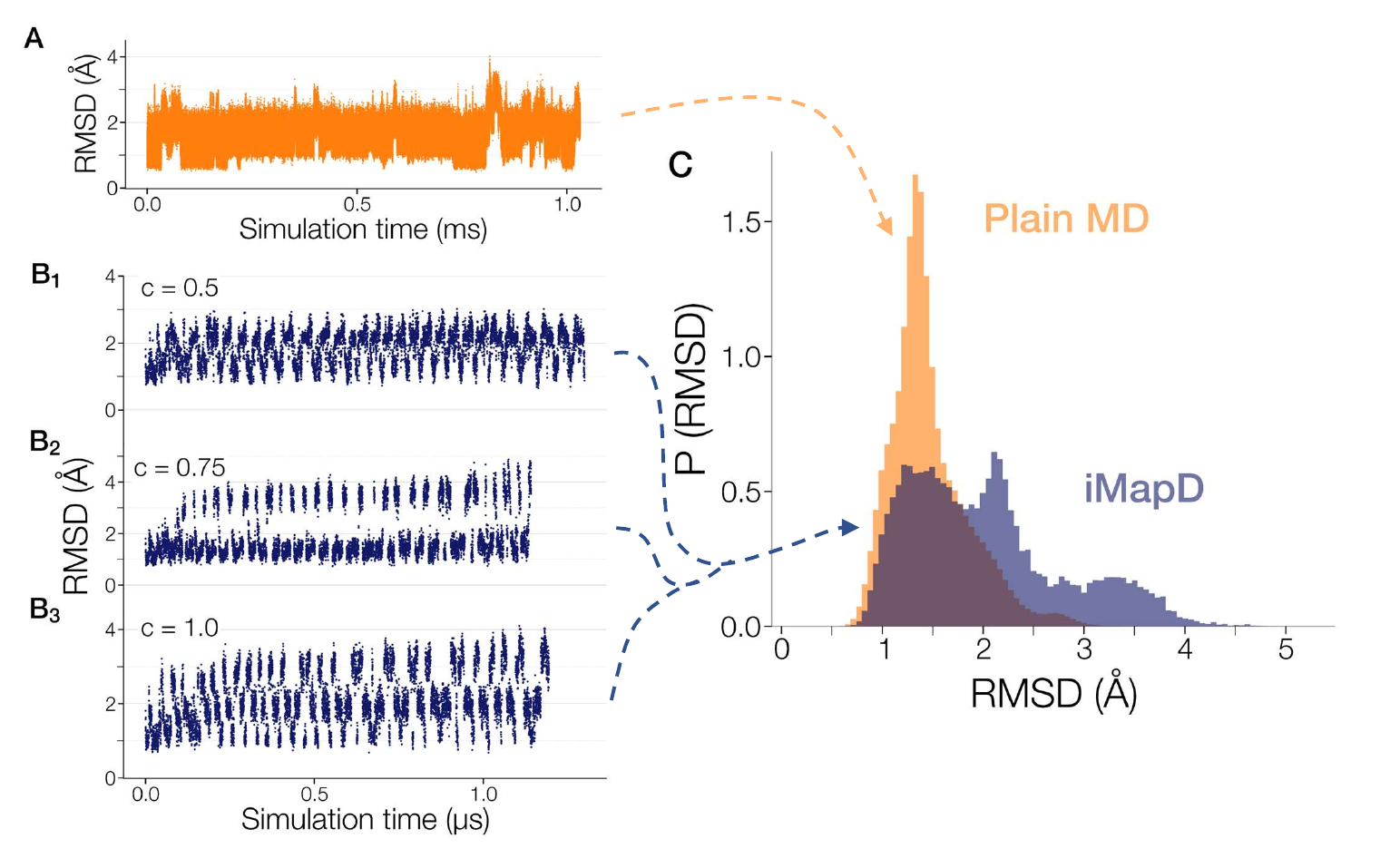}
\caption{ Panel \textbf{A}: the time series of the RMSD (evaluated on backbone atoms) to the first frame in Anton plain MD trajectory, taken as a definition of the native state.
 Panels \textbf{B}$_1$-\textbf{B}$_3$: the equivalent time series evaluated along the exploration cycles of the iMapD, with three different values of the parameter $c$ which controls the amplitude of the translation that drives the system outside the explored region (see definitions in Section S1 in \textrm{SI}). The oscillations visible in RMSD time-series of iMapD are associated with different exploration cycles. Panel \textbf{C}: The distributions of RMSD generated by plain MD and iMapD. }\label{fig_result1}
\end{figure}

Different meta-stable states visited by the plain MD trajectory are visible in panel A of Fig.~\ref{fig_result1}, where the molecular configurations are projected on a single collective variable, namely the RMSD from the first frame of the trajectory (taken as definition of native structure). These meta-stable states are reached by disrupting a few native hydrogen bonds in the small $\alpha$-helix, rotation of a aromatic rings, and by the chirality of a disulfide bond (see Section S5 in the \textrm{SI}).
In the panels B$_1$ through B$_3$ of Fig.~\ref{fig_result1}, we report the analog RMSD time series generated during the iMapD exploration, using the same all-atom force field (AMBERff99SB implemented in GROMACS software \cite{gromacs1,gromacs2}) in explicit TIP3P water.

These results show that, our improved version of iMapD yields trajectories that rapidly reach configurations with an RMSD distance from the initial structure comparable to that observed in the Anton plain MD trajectory.

In panel C, we report the cumulative distribution of RMSD evaluated from a frequency histogram of all the configurations in the plain MD and iMapD trajectories. This comparison highlights how iMapD does not yields the Boltzmann distribution at room temperature. This is expected since, even though each short MD trajectory generated in the iMapD exploration is fully unbiased and obeys the microscopic reversibility condition, the initial conditions are not drawn from the Boltzmann distribution, thus breaking the detailed balance condition.

This implies that the identification and characterization of the kinetically relevant metastable states cannot be carried out using standard dynamic clustering analyses such as those routinely adopted to define MSMs (see e.g. \cite{NoeMSM}). 
In section S5 in \textrm{SI}, we report an analysis based on a range of structural features and show that iMapD visits all the most populated states observed in the Anton simulation, missing only the most scarcely populated one. 
An interesting question to address is if the exploration could be performed with comparable accuracy by running plain MD simulations at high temperature. In Section S6 in \textrm{SI}, we compare the results based on iMapD with those obtained with plain MD at three high temperatures by projecting them on the plane selected by the RMSD distance to the native state and the fraction of native contacts $Q$.
After 300 ns of simulation times, none of the high-temperature MD runs covered the entire portion in this plane that is visited by the Anton MD trajectory. Conversely, iMapD extends significantly further than high temperature MD and covers the entire region explored by Anton, by traveling along a low free-energy valley. 

{\bf Sampling transition paths}. Next, we address whether the gTPS scheme correctly samples the transition pathways between these meta-stable states. According to the general gTPS protocol defined in the Theory section, we used a structural clustering algorithm to harvest from the iMapD simulations a subset of $\nu=80$ configurations uniformly distributed on the intrinsic manifold (i.e., with a nearly constant RMSD distance from the nearest neighbor), forming the structural data-set $\mathcal{C}$. In Section S2 of the \textrm{SI}, we compare different clustering schemes and motivate our choice of eventually relying on the hierarchical clustering with average linkage method (UPGMA) \cite{clustering_preprint_2011}, which ensures better coverage of the transition regions. 

In Fig. S7, the $\nu$ selected representative centroids (orange points) are projected onto the plane selected by two CVs and are superimposed to data-set generated by iMapD (blue points) and the free-energy landscape calculated from MD. 
The orange points correspond to the centers of the hyperspherical regions outlined in Fig.~\ref{fig_result1}B$_1$, with an average nearest-neighbor RMSD  $d\simeq 0.16 $ nm. We note that they cover the entire region in this Q-RMSD plane that the MD trajectory visits. However, their approximatively uniform density does not reflect the underlying Boltzmann distribution. 

To retrieve kinetic information and restore microscopic reversibility, the data-set $\mathcal{C}$ was used to build the network representation of the stochastic dynamics, according to the theory described in step 2 in Theory. The spatial resolution of this coarse-grained representation is estimated as half the average RMSD between neighboring configurations in $\mathcal{C}$, thus $\sigma\sim \sqrt{N} d/2 = \sqrt{N} 0.08$ nm, where $N$ is the number of (backbone) atoms in the protein. The time resolution of this coarse-grained theory is estimated from the average lifetime of the nodes in the network, i.e. $\Delta t= \langle \tau \rangle = s_0^{-1}\simeq 20$ ps. 

The $\nu=80$ configurations in the data-set $\mathcal{C}$ were then mapped onto the nodes of a graph, while the edges' weights $w_{ij}$ were calculated according to Eq.~(\ref{wijs0}). 
The coarse-grained effective potential $V^{\textrm{cg}}_{\textrm{eff}}(i_k)$ entering Eq.~(\ref{wijs0}) was calculated from lifetime $\tau$ of each node in the graph by running 100 separate simulations from each representative centroid configuration and then estimating the average time to reach a distance $\sigma$ from the initial centroid configuration (results of this calculations are shown in Fig. S2 in \textrm{SI}).

Finally, we resorted to the DWAVE quantum computer to efficiently sample transition pathways connecting two nodes on the network that lie very far in the Q-RMSD plane and for which we could detect at least one transition pathway in the Anton MD trajectory. We implemented the adiabatic switching procedure outlined in step 3 in the Theory section, using 207 qubits on the DWAVE quantum computer to perform our path sampling calculations. To look for low-action paths, we relied on the hybrid solver implemented on OCEAN (DWAVE's native programming suite) that integrates classical and quantum optimization.With this choice, the generation of a single transition path requires $\mathcal{O}(10^1)$~s of hybrid solver time (corresponding to $\mathcal{O}(1)$~s of QC time).

For comparison, we have also performed a number of analog calculations resorting to fully classical simulated annealing (as implemented in OCEAN). Even after increasing the sweeping number to $10^6$, which took $\mathcal{O}(10^3)$~s of computational time on a single core, most of the obtained paths have relatively low statistical weight: indeed, their action $S(\I)$ is typically two orders of magnitude larger than that of the least-action path calculated with a Dijkstra algorithm. However, a small portion of these paths (roughly $10\%$) has an action within a factor two of that of the least action path. In contrast,  the actions of all the paths generated with the hybrid quantum-classical are  within a factor 2 from that of the least action path (Fig. S7 in the \textrm{SI}.) 

\begin{figure*}[t!]
\centering
\includegraphics[width=1\textwidth]{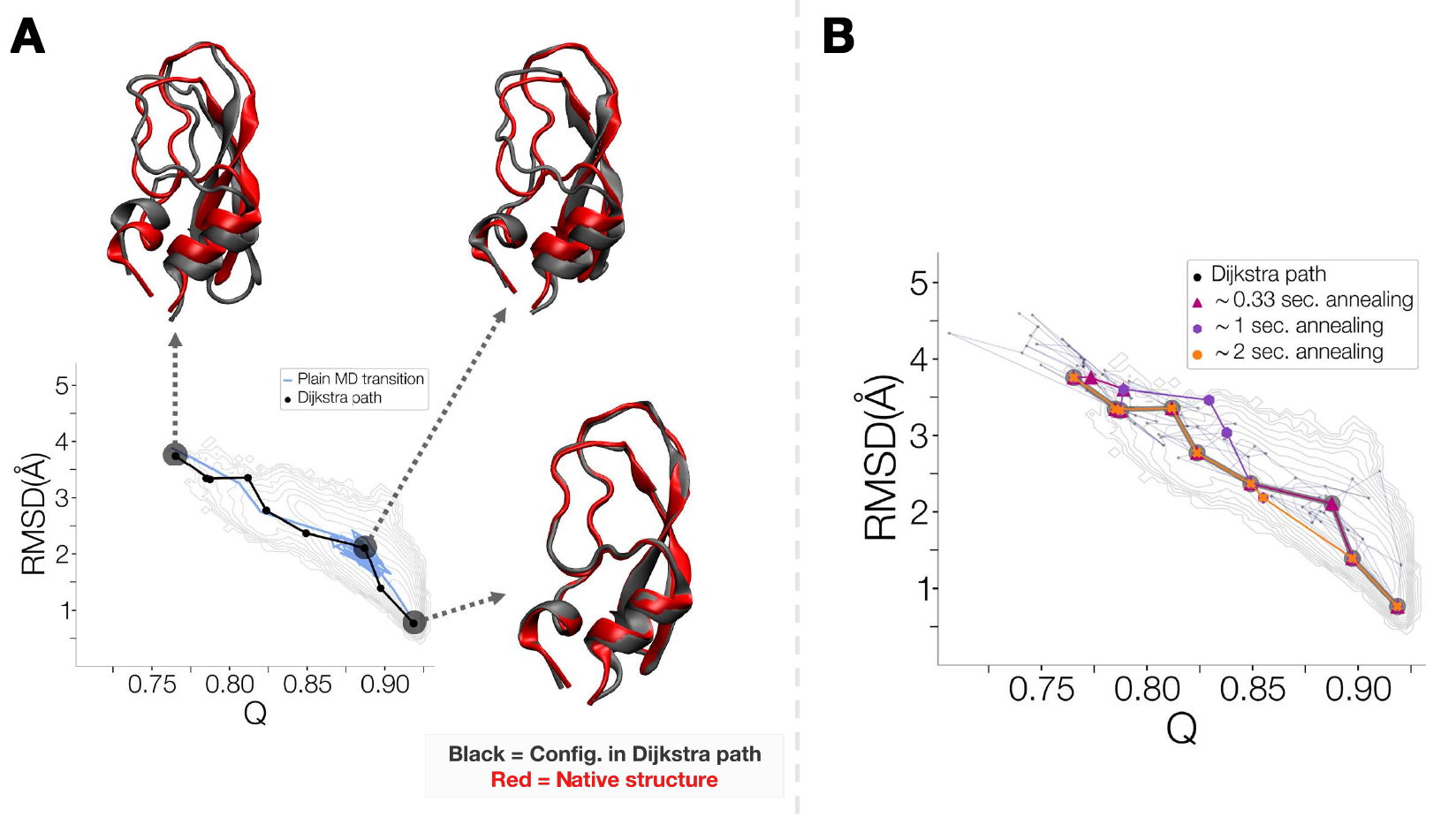}
\caption{Panel (\textbf{A}): The blue line follows the configurations in the transition pathway connecting two distant points in the plane selected by the RMSD to the native state and the fraction of native contacts, obtained by plain MD using the Anton supercomputer. Except for the beginning and the end of the line, the rest depicts the result of an average of over 50 ns-long windows. The solid black line is the DRP computed in the gTPS scheme, calculated using the Dijkstra algorithm. Selected representative conformations along the DRP are also included. To illustrate the evolution, these conformations (Black) are superposed onto the native structure (Red). Panel (\textbf{B}): Some representative gTPS transition pathways sampled using the DWAVE quantum computers and compared with the DRP path. The contour lines in the background of both panels highlight the structure of the free-energy landscape, obtained from a frequency histogram of the millisecond long MD simulation performed on the Anton supercomputer. }\label{fig_result3}
\end{figure*}

In Fig.~\ref{fig_result3}, we assess some of the paths generated by gTPS against the Anton plain MD simulations results. First, we compare the MD transition pathway observed in the Anton trajectory with the most probable transition path on our network initiated from the configuration closest to the initial frame of the Anton productive transition path. 
This representative path (also known as Dominant Reaction Pathway (DRP) \cite{DRP1, DRP2}) was calculated on a classical machine using the Dijkstra method. Alternatively, the implementation on a quantum annealing machine is also quite efficient, as shown in Ref. \cite{QDRP}. 

In panel B of Fig.~\ref{fig_result3}, we report some representative stochastic trajectories obtained by quantum computing, using the DWAVE quantum computer. Additional trajectories in the \textrm{SI} are available in Fig. S8, but have been omitted here for visual clarity. 
 
As expected, the Anton's MD trajectory, the DRP, and the gTPS stochastic trajectories evaluated using DWAVE reach the final destination by traveling along a low-free energy region. This result assesses the accuracy of the gTPS in producing viable reactive pathways at an affordable computational cost. 

A significant fraction of the frames of the MD transition accumulates in the vicinity of one of the frames, indicating the presence of a meta-stable state. As discussed in the Theory section, the DRP and the stochastic paths obtained by gTPS miss this information, as they ignore the exploration of the metastable states. 

 {\bf Transition path time.} The gTPS formalism yields a lower-bound of the transition path time for transitions that overcome a single energy barrier. 
For the specific case of the BPTI transition shown in panel A of Fig.\ref{fig_result3}, using Eq.~(S14) of the \textrm{SI} on the least-action path, we obtain a lower-bound estimate $t\sim \,500$~ps.
This time scale cannot be directly compared with the transition path time observed in the corresponding plain MD productive trajectory, since the latter involves overcoming more than one energy barrier. 
However, in Fig. S3 of the \textrm{SI}, we analyze the MD transition pathway by projecting it onto the RMSD to the native state and isolating the sections involving a transition between metastable basins, finding $t\sim\,1.2$~ns. Hence, the transition time observed in MD is about a factor of 2 longer than the lower bound estimate provided by gTPS. 

\section{Methods}
In application of iMapD to BPTI, we initially solvated the crystal structure of the molecule in a square box of size $6$-nm with $6744$ molecules of TIP3P water using the forcefield AMBER99SB-ILDN, implemented in GROMACS \cite{gromacs1,gromacs2}. The system was neutralized with 6 ions of chloride. Next, we performed energy minimization and equilibrated the system at $300$ K in the NVT ensemble for 200-ps with an integration step of 1~fs. We resorted to a stochastic velocity rescaling thermostat, with coupling constant $0.1\,\text{ps}^{-1}$. Finally, we simulated for $10$-ns in the same NVT ensemble. Configurations were saved every $100$-ps.

The exploration of the protein energy landscape is made challenging by the co-existence of structures at different length scales. To cope with this issue, we integrated the results of iMapD simulations performed with 3 different values of "shooting strengths" $c=0.5,\,0.75,\,1.0$ defined in \textrm{SI}. The calculation of the contact map were carried out using MDTraj package in python \cite{mdtraj}. To perform rMD simulations, we relied on an in-house modified version of GROMACS. Equivalently, this calculation could be performed using the publicly available PLUMED plugin\cite{PLUMED}.

For each choice of $c$, we performed 30 iMapD exploration cycles. In each cycle, 5 individual $1$-ns long MD simulations were performed starting from configurations outside the boundary generated by the Polar Star. In each exploration cycle, the scaling parameter that defines the DMAP kernel was set to the average value of pairwise-RMSD between all the configurations sampled by iMapD up to that cycle.

To analyze molecular configurations, extract features and compute CVs we resorted to MDTraj. The fraction of contacts $Q$ calculated relatively to a given reference was computed according to
\begin{equation}
 Q=\sum_{i<j\;(\,\in\textrm{C}\,)}\frac{1}{1+e^{\beta (d_{ij} - l_a r_0)}},
\end{equation}
where $\beta=5$ and $l_a=1.5$. The summation runs over the set of indexes $\textrm{C}$ of all the residues that are in contact in the reference configuration, based on the $C_\alpha$ distance criterion $d_{ij}\leq r_0$ with $|j-i|>3$, $r_0=7.5\AA$.

\section{Discussion}

In this work, we reported our first application of the gTPS scheme to investigate a rare protein structural rearrangement that spontaneously occurs on a millisecond time scale, using an all-atom force field in explicit solvent.

We introduced an improved version of the iMapD algorithm \cite{IMAPD} to rapidly explore relevant macromolecular configurations near the native state of the BTPI protein. By using the Polar Star shooting scheme, we could generate configurations that were kinetically distant from the starting native one.

We used these configurations to define a graph representation of the kinetics consistent with microscopic reversibility. We emphasize that this coarse-grained description is different from that of MSMs: It is based on an undirected graph and, most importantly, it does not require estimating the whole stochastic transition matrix (or rate matrix). Instead, it only requires computing the life-time of each node, a task that can be efficiently carried out even on-the-flight, during the uncharted exploration. On the other hand, it can only provide limited information about the kinetics, neglecting the time spent exploring metastable states. 

The transition paths obtained from the DWAVE quantum computer are consistent with the free-energy landscape estimated from a frequency histogram of the ultra-long Anton MD trajectory, projected on a suitable plane defined by a standard pair of CVs. 

Applications of quantum computing technologies to computational biophysics and soft-condensed matter systems have insofar mostly relied on coarse-grained lattice-based polymer representations \cite{QPFOLDING1, QPFOLDING2, QPFOLDING3, ASPURU, POLYQ}. 
This work shows that quantum computers can be successfully applied to investigate challenging protein transitions with full atomic resolution and in explicit solvent. In particular, the DWAVE quantum computer can generate viable transition pathways at an affordable computational cost (with a calculation time of a few seconds per path, using the OCEAN's hybrid solver). Most importantly, we have previously shown that the initialization of the quantum computer's wave function in an equal superposition of qubit states ensures that all the paths obtained at the end of the annealing process are completely uncorrelated \cite{QTPS}. 

The limitations in size and performance of the existing quantum annealing machines forced us to represent the kinetics using a relatively small network. On such a small network, it is hard to assess the potential of the gTPS scheme for reaching a quantum advantage relative to alternative path sampling schemes based on classical computing, e.g. via kinetic Monte Carlo. On the other hand, previous studies have shown that quantum annealers are in fact quite efficient in generating transition paths \cite{QDRP} or equilibrium polymer configurations \cite{POLYQ}. If quantum computers continue growing exponentially in size and performance, the approach introduced in this work may provide in the foreseeable future a computationally efficient scheme to perform path sampling calculations for complex macromolecular transitions. 

A very important quest to be pursued in this phase of the development of quantum technologies is identifying possible new fields of applicability. These results identify all-atom macromolecular simulations as a new field where quantum technologies can play an important role. Conversely, biomolecular simulations can guide the development of quantum hardware and inspire new quantum algorithms. 
\section*{Acknowledgments}
We are thankful to C.Micheletti and P. Hauke for useful discussions. R.C. acknowledge the support of the Frankfurt Institute of Advanced Studies, the LOEWE Center for Multiscale Modelling in Life Sciences of the state of Hesse, the CRC 1507: Membrane-associated Protein Assemblies, Machineries, and Supercomplexes, and the International Max Planck Research School on Cellular Biophysics. 

\section*{Conflict of Interest}
PF is a co-founder and shareholder of Sibylla Biotech SPA, a company involved in exploiting molecular simulations to perform early-stage drug discovery. 
\setcounter{section}{0}
\setcounter{table}{0}
\setcounter{figure}{0}
\setcounter{equation}{0}

\renewcommand{\thesection}{S\arabic{section}} 
\renewcommand{\thetable}{S\arabic{table}} 
\renewcommand{\thefigure}{S\arabic{figure}}
\renewcommand{\theequation}{S\arabic{equation}}
\section*{Supplementary Information}
In this Supplementary Information, we provide details about the mathematical formalism and the numerical implementation of the graph Transition Path Sampling (gTPS) algorithm, and its quantum encoding based on Quadratic Unconstrained Binary Optimization (QUBO). We report numerical results concerning the application of this scheme to investigate the native dynamics of Bovine basic Pancreatic Trypsin Inhibitor (BPTI) protein, and a detailed comparison with the results of the plain MD simulations performed on the Anton supercomputer. Moreover, we provide additional comparison between the results of data-driven exploration based on Intrinsic Map Dynamics (iMapD) algorithm~\cite{IMAPD} and the one with high-temperature MD.

\section{Uncharted manifold exploration}
This section briefly reviews the {uncharted exploration framework of iMapD algorithm.}
We also provide details about the Polar Star scheme, which we introduced to accelerate the iMapD exploration. 
 
\subsection{The iMapD approach}
\label{section_IMAPD}
The first step of iMapD consists of performing a short Molecular Dynamics (MD) simulation, starting from some given initial configuration. In the specific application to 
the protein BPTI, we initiated the exploration from the energy-minimized crystallographic structure. The $m$ resulting configurations are then projected onto a low-dimensional space, using the Diffusion Map (DMAP) formalism \cite{DMAP}, {for which in the present work we only retained the first two nontrivial components}

{In the next step we identify} the set of configurations that, in the projected space, form a convex hull surrounding the initial $m$ configurations. This procedure enables the definition of a boundary of the explored region.

We then resort to Principal Component Analysis (PCA) to learn the local structure of the dataset in the neighborhood of each boundary point, $X_i^b$. {Here after forming the set of neighboring configurations $\mathcal{B}_i$, we perform PCA on this set to find the loading matrix $\mathcal{L}_i$ which retains a relevant subset of $P$ dominant PCA components, saturating $98\%$ of the sum of all PCA eigenvalues. In the next step, we project the average Cartesian coordinates of the configurations in $\mathcal{B}_i$, denoted as $X^{{\mathrm{avg}}}_i$, and $X_i^b$ using
\begin{equation}
 x^b_i = \mathcal{L}_iX^b_i,\quad x^{\mathrm{avg}}_i = \mathcal{L}_iX^{\mathrm{avg}}_i
\end{equation}
}
%

To generate a new unexplored configuration $X'_i$ beyond the boundary, we perform a translation in the PCA dominant subspace:
\begin{equation}\label{translation}
 y_i = x_{i}^b-x_{{\mathrm{avg}}}^{(i)}+c\frac{x_{i}^b-x_{{\mathrm{avg}}}^{(i)}}{|x_{i}^b-x_{{\mathrm{avg}}}^{(i)}|}, 
\end{equation}
where $c$ is a tunable parameter that defines the amplitude of the shooting move. 
We note that $(x_{z}^b-x_{{\mathrm{avg}}}^{(i)})/|x_{i}^b-x_{{\mathrm{avg}}}^{(i)}|$ provides an approximation to the direction perpendicular to the surface of the boundary of the explored region, in the point $x_i^b$. 
Finally, we perform the reverse transformation $Y_i=y_i\mathcal{L}^T+X_{{\mathrm{avg}}}^b$ to embed these points in configuration space. The corresponding set of Cartesian coordinates lie outside of the boundary of intrinsic manifold. As mentioned in the main text, the new configurations $Y_i$ generated this way do not represent viable molecular configurations, as they do not respect the topological constraints dictated by the protein chemical structure and by steric avoidance. However in principle, a viable molecular configuration $X_i'$ outside the boundary may be obtained by energy minimization starting from $Y_i$.

The new configurations $X_i'$ outside the border of the explored region are used as starting points in further iteration cycles of iMapD algorithm. After each iteration, the new configurations are appended to the dataset defining the explored region.

\subsection{Limitations of the conventional iMapD and the Polar Star scheme}

As discussed above, in the original version of iMapD, a viable molecular configuration outside the border of the explored region is obtained by energy minimization, starting from the set of Cartesian coordinates generated by the translation (\ref{translation}). Unfortunately, this procedure is numerically stable only for incremental translations, i.e for very small values of the parameter $c$. As a result, the exploration of complex systems can be slow and computationally challenging.
As discussed in the main text, we introduced a scheme called Polar Star to overcome this limitation,. 
This is based on replacing the energy minimization step with a short run of a specific type of biased dynamics called ratchet-and-pawl-MD (rMD)\cite{rMD1,rMD2}. 

In a rMD simulations, the equations of motion of the macromolecular system are modified by introducing a history-dependent biasing force defined as:
\begin{align}
 {\bf F}^i_B&(X, q_m(t) ) =\nonumber \\
 &- k_\textrm{rMD} \nabla_i q(X) (q(X)-q_\textrm{m}(t)) \theta[q(X)-q_m(t)]
\end{align}
where $\theta(X)$ is the Heaviside step function, $X=({\bf x}_1, \ldots,{\bf x}_N)$ denotes the set of atomic coordinates and $q(X)$ is a Collective Variable (CV), while $q_m(t)$ denotes the maximum value attained by $q(X)$ up to time $t$. Note that the dynamics is unbiased unless the system tends to evolve to a configuration with a larger value of $q(X)$.

In our present application,
the CV $q$ measures the instantaneous overlap with the a target contact map $C^0_{ij}$:
 \begin{equation}\label{qdef}
 \tilde{q}(X) = \frac{\sum_{|i-j|> 35} (C_{ij}(X)-C^0_{ij})^2}{\sum'_{i,j}C^{0\, 2}_{ij}},
 \end{equation}
In this equation, $C_{ij}(X)$ is a switching function that approaches 1 when atom $i$ and $j$ are in contact and vanishes when they are far apart. In particular, we use: 
\begin{eqnarray}\label{Cij}
C_{ij}(X) = \frac{1- \left(\frac{r_{ij}}{r_0}\right)^6}{1- \left(\frac{r_{ij}}{r_0}\right)^{10}}.
\label{contact_map_eq}
\end{eqnarray}
Here, $r_{ij}= |{\bf x}_i - {\bf x}_j|$ and $r_0=4.5 $~\AA~is a threshold reference distance and $C_{ij}^0$ are entries of the contact map of the target state. We note the restriction $\sum_{|i-j|> 35}$ in Eq. (\ref{qdef}), which excludes from the summation pairs of atoms that belong to neighboring amino acids. This restriction is introduced to avoid the biasing force to act on atom pairs which are subject to strong correlations determined by the local chemical structure of the chain.

In the application of rMD to Polar Star shooting, the target contact map $C^0_{ij}$ is defined in a data-driven way: 
\begin{equation}
C^0_{ij} = C_{ij}(Y_i)
\end{equation}
where $Y_i$ is the set of Cartesian coordinates of a point outside the boundary of the explored region, generated with Eq. (\ref{translation}). An rMD simulation initiated from the boundary point $X_i^b$, rapidly yields a new viable molecular configuration outside the boundary, with a contact map very close to $C^0_{ij}$. The name "Polar Star" is based on the analogy with the ancient navigation scheme based on pointing towards a star (outside the Earth's surface manifold) to orient the sailing direction. 

\section{Details about the gTPS network definition}

\subsection{Clustering}\label{clustering_section}
Following the gTPS framework, once the iMapD data is generated, we select configurations which uniformly (or semi-uniformly) cover the explored region of configuration space. 

In this work, we have applied and compared 3 different methods of structural clustering \cite{kmeans_paper,clustering_preprint_2011}: KMeans, Hierarchical clustering (HC) with unweighted average linkage, and HC with Centroid linkage. The best clustering for this task was established using two independent criteria: i) the distribution of nearest-neighbor RMSD calculated over the centroids of the clusters should be narrowest, denoting a uniform spacing of the clusters, and ii) The centroids of the clusters should be uniformly distributed in the plane defined by $Q$ and the RMSD distance to the native structure (defined as the first configuration in the Anton MD trajectory). 

{The KMeans approach aims to cluster the data while minimizing the average intra-cluster Euclidean distance between members of each cluster. In our application, KMeans} was performed using the scikit-learn \cite{scikit-learn} package with $C_\alpha$-atoms' coordinates taken as the main initial features of the data set. We set the algorithm to find the optimum clustering after 20 runs of randomly initializing the centroid of the clusters using the "\textit{kmeans++}" built-in method. The algorithm was then terminated once in successive iterations the distance between the consecutive cluster centers fell below $0.1\AA$. 

Alternatively, in HC starting by taking every point as an isolated cluster, at each step we form new clusters by connecting those with the minimum "linkage" distance in the previous step. For HC with unweighted average linkage --UPGMA method--, this distance is defined as
\begin{equation}
 D(A,B) = \frac{1}{n_An_B}\sum_{{X_a}\in A}\sum_{{X_b}\in B}d({X_a},{X_b})
\end{equation}
where, in our application, the $d({X_a},{X_b})$ is the RMSD of the backbone atoms, and $n_\text{K}$ is the number of elements in cluster $\text{K}$. 
Conversely, in Centroid linkage, the Euclidean distance between the center of centroids are taken as the dissimilarity (distance) between clusters
\begin{equation}
 D(A,B)=|{\mu_A}-{\mu_B}|
\end{equation}
where $\mu_K=(1/n_K)\sum_{X_k\in K}{X_k}$ denotes the centroids' center.

\begin{figure}\centering
 \includegraphics[width=1 \textwidth]{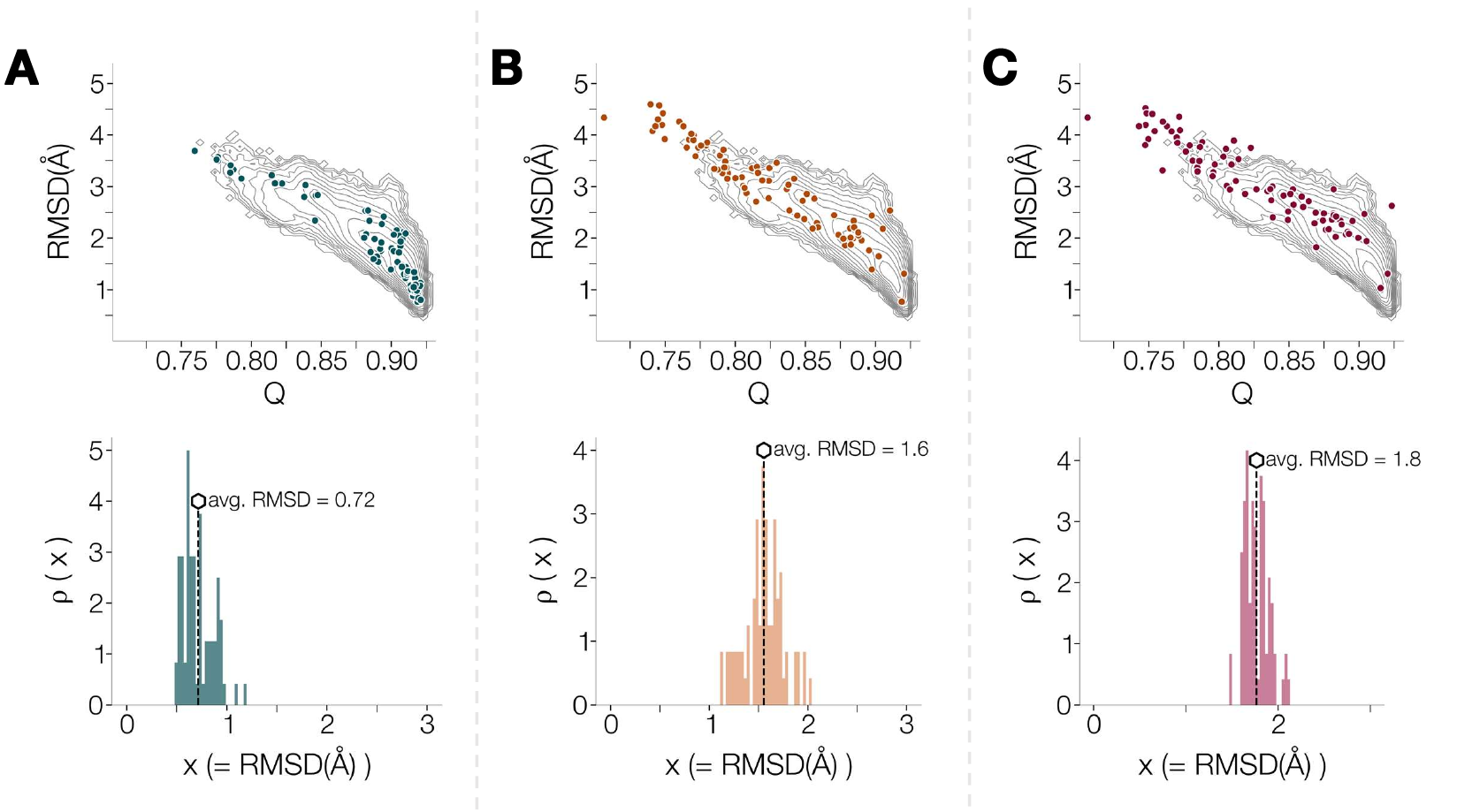}
 \caption{Results of clustering iMapD configurations using KMeans (panel A), HC with Average linkage (panel B), and HC with Centroid linkage (panel C). In all panels, Q-RMSD plots on the top depict the representative configurations characterized by the least average {intra-cluster RMSD.}
 The figures at the bottom report the associated distributions of nearest neighbor distances between configurations in different clusters. }
 \label{SI_clustering_fig}
\end{figure}

For both HC approaches, we relied on the SciPy package \cite{scipy} with default values for all the parameters.
For each clustering method, we selected a representative element of each cluster as the configuration with the least average pairwise-RMSD from all other members of that cluster. The results are reported in Fig. \ref{SI_clustering_fig}. Each panel shows the representative configurations in the $Q$-RMSD plane (top) and the distribution of nearest neighbors (bottom). Based on this set of results, we identified HC-with Average linkage as the algorithm that better meets the two criteria above.

\subsection{Construction of the gTPS graph}\label{network_section}

In defining our network representation of the dynamics, first the representative cluster elements were assigned to the nodes of the graph, while the edges were assigned to nodes with a relative pairwise RMSD $\leq d_{\textrm{cutoff}}$, where $d_{\textrm{cutoff}}\,=\,1.6\AA$. Next we established connections from every vertex with degree $<2$ to utmost 2 of their nearest neighbors --albeit with RMSD$\,>\, d_{\textrm{cutoff}}$. This step is essential to allow the possibility for the paths to visit every node. Finally to ensure the topological connectivity of the graph, we linked each disconnected component to its nearest neighbor using their vertices with minimum RMSD. 

Once the network was established, we calculated the weights of each edge using Eq. (3) of the main text. 

\section{gTPS formalism and the transition path time. }

Here, we briefly review the main equations of the gTPS formalism and use them to obtain our lower-bound estimate of the transition path time. 

\begin{figure}[t!]\centering
 \includegraphics[width=1\textwidth]{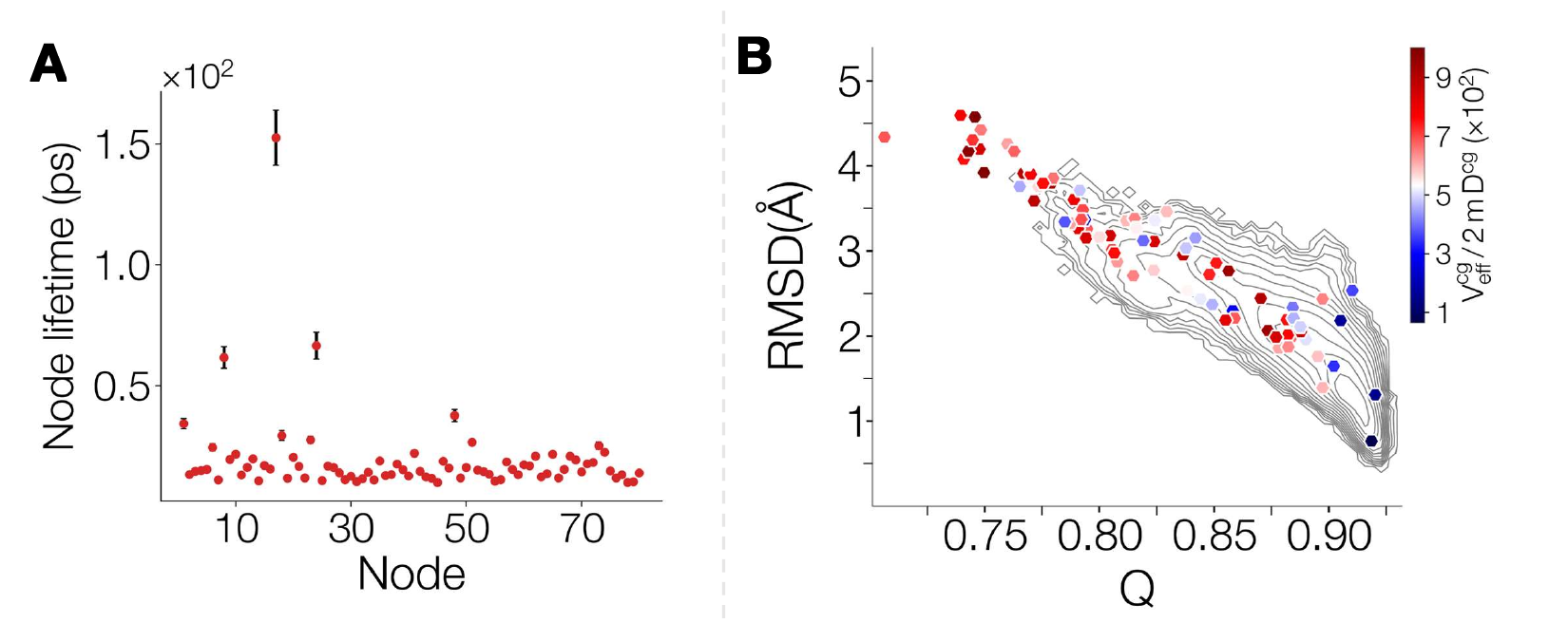}
 \caption{The results of the calculation of the coarse-grained effective potential $\text{V}_{\textrm{eff}}^{\textrm{cg}}$ associated to each node of the graph (i.e. the clustered data shown in Fig. \ref{SI_clustering_fig}). $\text{V}_{\textrm{eff}}^{\textrm{cg}}$ was calculated as the inverse of node life-time (reported in panel A). Panel B shows the value of $\text{V}_{\textrm{eff}}^{\textrm{cg}}$ in the Q-RMSD plane.}
 \label{Vcg}
\end{figure}
The gTPS theory follows from systematically lowering the spatial and temporal resolution of the underlying Langevin dynamics. To estimate the probability of a given path in the gTPS network and the corresponding time duration, we rely on the so-called Dominant Reaction Pathway (DRP) formalism introduced in Ref.s \cite{DRP1, DRP2}. 

According to the DRP theory,
the most probable transition pathway connecting two points $X_i$ and $X_f$ in configuration space is obtained by minimizing the so-called Hamilton-Jacobi functional:
\begin{equation}\label{SHJ}
S_{HJ}[X] = \int_{X_i}^{X_f} dl \sqrt{\frac{1}{D}\left(\frac{\text{V}_{\textrm{eff}}[X(l)]}{2 m D}+ s_0\right)}.
\end{equation} 
Here, $D$ is the diffusion coefficient, $m$ is the mass, and $l$ is a curvilinear abscissa that measures the Euclidean distance covered by the reactive trajectory $X(l)$. $\text{V}_\textrm{eff}(X)$ is the effective potential, defined as:
\begin{equation}
 \text{V}_\textrm{eff}(X)= \frac{m D^2}{ 2 (k_B T)^2 }\left(|\nabla U(X)|^2 -2 k_BT \nabla^2 U(X)\right),
\end{equation}
where $U(X)$ is the potential energy. 

$s_0$ is a frequency scale, which is related to the time taken by the most probable path $\bar{X}(l)$ to connect the endpoints $X_i$ and $X_f$:
\begin{eqnarray}\label{time}
 t= \int_{X_i}^{X_f} dl \frac{1}{\sqrt{4 D \left( \frac{\text{V}_{\textrm{eff}}((\bar{X}(l))}{2 m D}+ s_0\right)}}.
\end{eqnarray}
 
In the gTPS theory, Eq.s (\ref{SHJ}) and (\ref{time}) are replaced by: 
\begin{eqnarray}
S_{HJ} &=& \sum_{i=1}^{N-1} \Delta l_{i i+1}\sqrt{\frac{1}{D_{cg}} \left(\frac{\text{V}_{\textrm{eff}}^{\textrm{cg}}(i)}{2 m D_{cg}}+ s_0\right)} \\
 t &\simeq& \sum_{i=1}^{N-1}
 \Delta l_{i i+1} \frac{1}{\sqrt{4 D_{cg} \left(\frac{\text{V}_{\textrm{eff}}^{\textrm{cg}}(i)}{2 m D_{cg}}+ s_0\right)}}.
\end{eqnarray}
Here, the summation runs over the nodes visited in order, by a given path, $\Delta l_{i i+1}$ is the distance between two consecutive visited nodes, $m$ is an average atomic mass. $D_{cg}= \frac{\sigma^2}{2 \Delta t}$ is the coarse-grained diffusion coefficient, which is defined in terms of the gTPS spatial and temporal native resolution scales. The former corresponds to half the average distance between neighboring nodes in the graph, $ \sigma = \frac{1}{2}\langle \Delta l\rangle $, while the latter is the average lifetime of the nodes in the graph $\Delta t= \langle \tau \rangle$. Finally, $s_0=1/\Delta t$ is the associated frequency cut-off scale.
$\text{V}_{\textrm{eff}}^{\textrm{cg}}(i)$ is the 
coarse-grained effective potential at the $i$-th node, which is directly related to the node's inverse lifetime: 
\begin{equation}
 \tau_i = \frac{2m D_{cg}}{ \text{V}_{\textrm{eff}}^{\textrm{cg}}(i)}
\end{equation}

In Fig. \ref{Vcg}, we report the values of effective potential calculated from short MD simulations initiated from the centroids associated to each node in the gTPS graph for protein BPTI.

We now approximate all the distances $\Delta l_{i i+1}$ with their average value, $\langle \Delta l \rangle = 2 \sigma$ and substitute the definition of $D_{cg}$ and $s_0$ finding 
\begin{eqnarray}
t &\simeq& \sum_{i=1}^{N-1}
\frac{1}{\sqrt{\frac{1}{2\langle \tau \rangle} 
\left(
\frac{1}{\tau_i}+ \frac{1}{\langle \tau \rangle }
\right)}}\\
& \simeq& \sum_{i=1}^{N-1}
\frac{\langle \tau \rangle}{\sqrt{\frac{1}{2} 
\left(
\frac{\tau_i +\langle \tau \rangle}{\tau_i }
\right)}}
\end{eqnarray}

Finally, if we approximate all life times with their average value, we obtain our final simple expression: 
\begin{equation}
t \simeq N \langle \tau \rangle,
\end{equation}
where $N$ is the number of nodes visited by the transition path. 

\begin{figure}[t!]
\centering
\includegraphics[width=1 \textwidth]{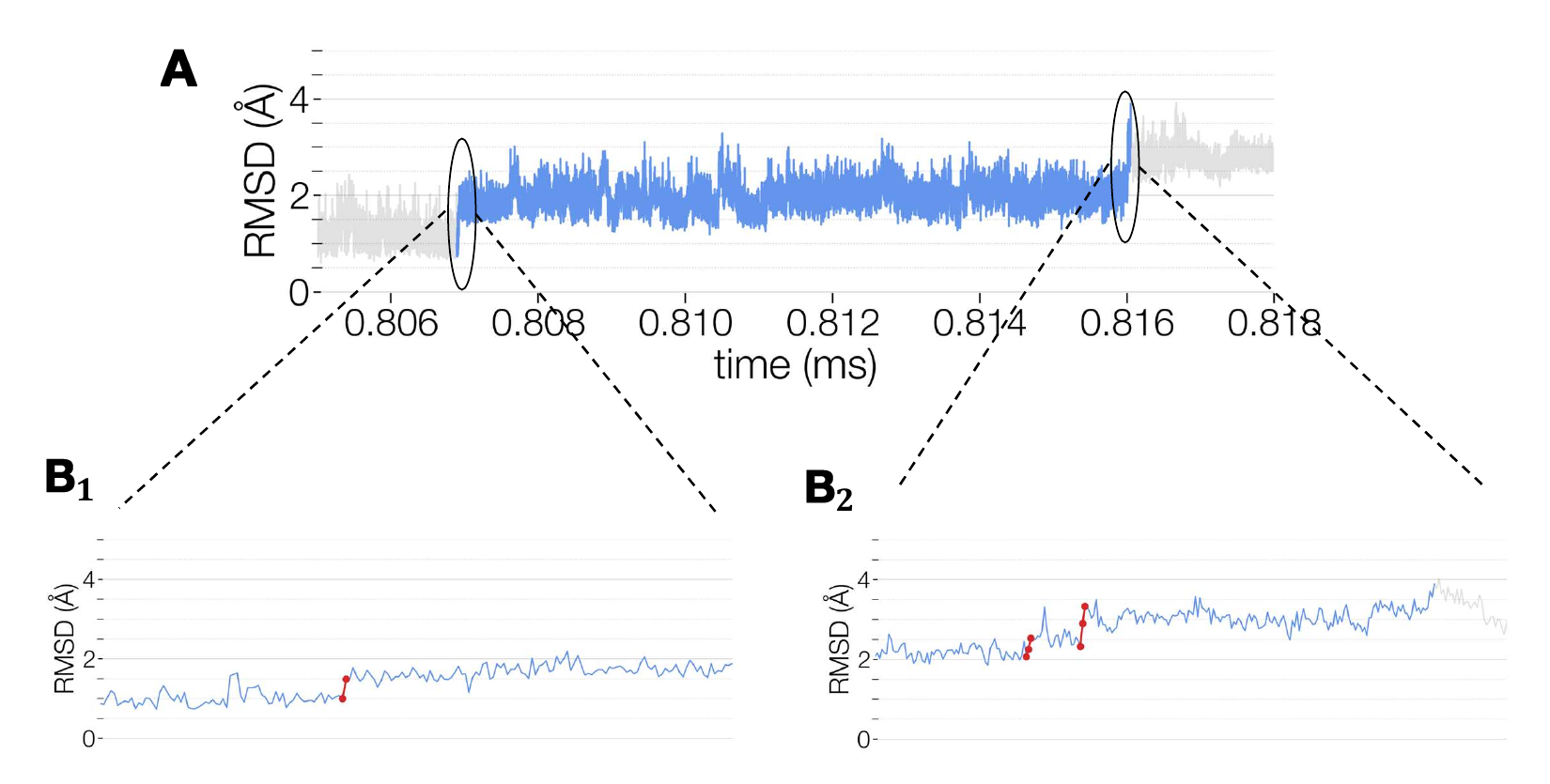}
\caption{MD transition path for the conformational transition of BPTI shown in Fig.5 of the main text, projected on the RMSD from the first frame. The 8 red points have been used to estimate the barrier-crossing time scale $t$ discussed in the Results section of the main text. }
\label{tMD}
\end{figure}

The frames in the MD Anton trajectory used to estimate $t$ for a conformational transition of protein BPTI are highlighted in Fig. \ref{tMD}.

\section{Details about the QUBO encoding of the path sampling problem}\label{QC_section}
The Hamiltonian $\hat{H}_\textrm{i}$ that used to prepare the initial state of the quantum annealing machine is defined as 
\begin{eqnarray}
 \hat H_\textrm{i} = - h\left( \sum_i \hat \sigma_x^i + \sum_{ij} \hat \sigma_x^{ij}\right), 
\end{eqnarray}
where $\hat \sigma_x^i$ and $\hat \sigma_x^{ij}$ are Pauli "spin-x" operators acting on the qubit associated to the $i-$th node and the $ij$ vertex, respectively and $h$ is an arbitrary real constant.

The constraint Hamiltonian $\hat H_\textrm{f}$, reached at the end of the adiabatic switching process is 
\begin{eqnarray}
\label{HQUBO}
\hat H_\textrm{f} = \alpha~ \hat H_{\textrm{C}} + \hat H_{\textrm{T}}.
\end{eqnarray} 
$H_{\textrm{C}}$ is the so-called constraint Hamiltonian, which ensures that all the low-lying eigenstates of $H_\textrm{f}$ obey path topology~\cite{shortest_path_on_network}. 
$\hat H_{\textrm{T}}$ is the so-called target Hamiltonian, which breaks the degeneracy of the ground state of $\hat H_{\textrm{C}}$ and orders the path states according to their relative probability in our coarse-grained theory. 

The constraint Hamiltonian reads
\begin{eqnarray}
 H_{\textrm{C}}= H_{\textrm{s}}+H_{\textrm{t}}+H_\textrm{r},
\end{eqnarray}
where 
\begin{eqnarray}
\hat H_{\textrm{s}}=& -\left(\hat \Gamma_s\right)^2+\left(\hat \Gamma_s-\sum_i\hat \Gamma_{s i}\right)^2\,,\\
\hat H_{\textrm{t}}=&- \left(\hat \Gamma_t\right)^2+\left(\hat \Gamma_t-\sum_i\hat \hat \Gamma_{t i}\right)^2\,,\\
\hat H_{\textrm{r}}=& \sum_{j\ne s,t} \left(2\hat \Gamma_j-\sum_i\hat \Gamma_{j i}\right)^2\,.
\end{eqnarray}
where $\hat \Gamma_i$ and $\hat \Gamma_{ij}$ are related to the spin-z operators that act on the qubits associated to the nodes and the vertexes of the graph, respectively: 
\begin{eqnarray}
 \hat\Gamma_i = \frac{\hat \sigma_z^i +1}{2}\\
 \hat \Gamma_{ij} = \frac{\hat \sigma_z^{ij} +1}{2}\\\nonumber
\end{eqnarray}
We note that the $\hat H_\textrm{s}$ and $\hat H_\textrm{t}$ Hamiltonians introduce the condition that the path should start from the initial node $s$ and end in the final node $t$, while $\hat H_\textrm{r}$ imposes the flux conservation at the remaining nodes. 

The target Hamiltonian is defined as: 
\begin{eqnarray}
 \hat H_\textrm{T}= \sum_{ij} w_{ij}~\hat \Gamma_{ij}
\end{eqnarray}
By definition, when $\hat H_\textrm{T}$ acts on a state in the ground-state of the constraint Hamiltonian, it yields an energy equal to the path action $S$ defined in Eq. (1) of the main text. 

The parameter $\alpha$ in Eq.~(\ref{HQUBO}) controls the relative strength of the constraint Hamiltonian, $H_\textrm{C}$.In order
for $\hat H_\textrm{C}$ to be a hard constraint, we require $\alpha$ to be sufficiently large. A rigorous
condition is $\alpha\gg \epsilon W_\textrm{max}$, $\epsilon$ is the number of edges, and $ W_\textrm{max}$ is the larger weight in the graph. In practice, however, smaller values often suffice.

\section{Comparison between the structures explored with iMapD and the metastable states found in the Anton MD trajectory}\label{state_indetification}

\begin{figure}[!t]\centering
 \includegraphics[width=1\textwidth]{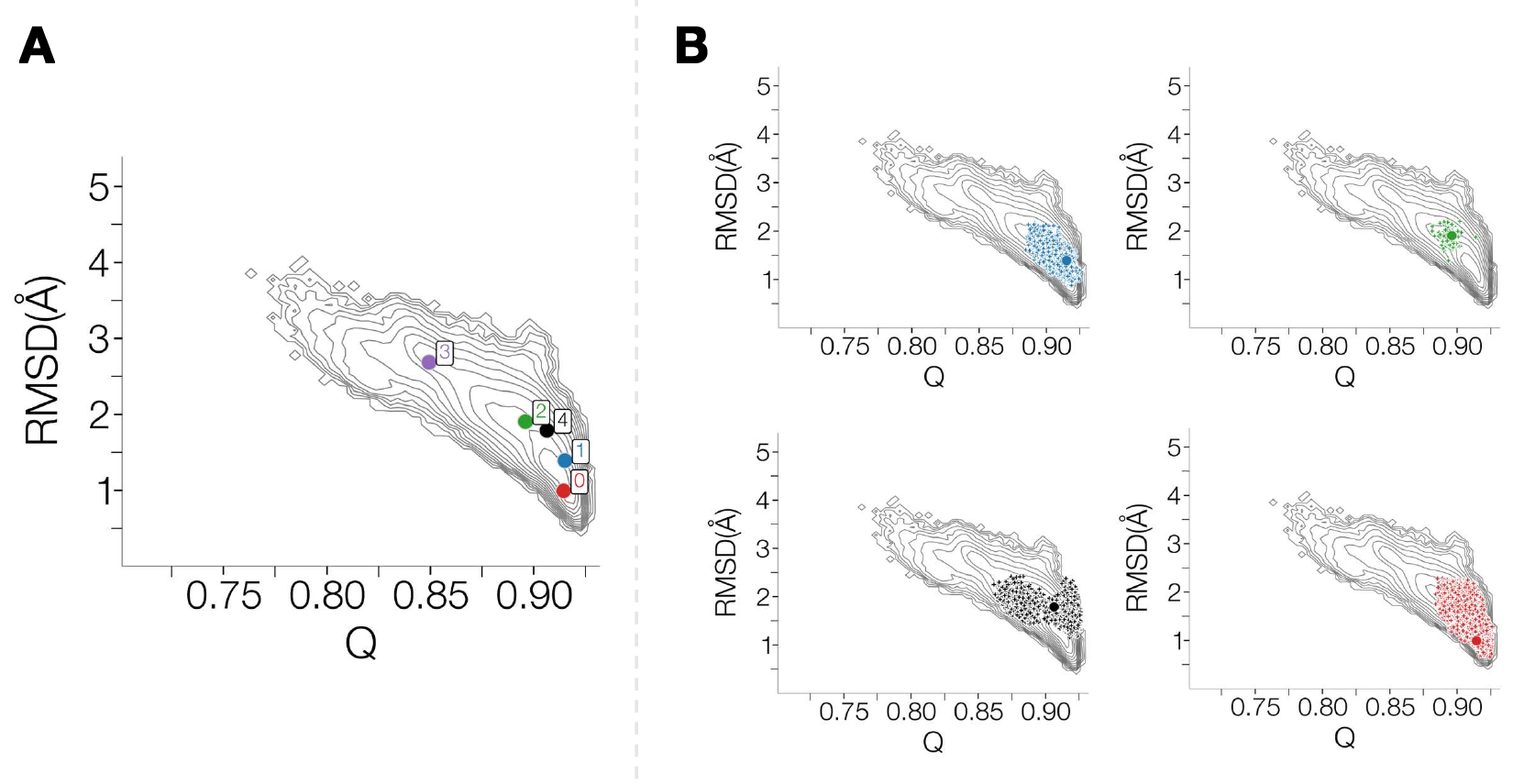}
 \caption{Configurations in iMapD data that were identified as corresponding to Anton states, using the Q$>0.9$ and pairwise-RMSD $<2\AA$ as outlined in section \ref{state_indetification}. Using this criteria, no configuration was associated with the state-3 of Anton. Color coding is the same as in original article \cite{Anton2010}.}\label{SI_states_anton_fig}
\end{figure}

The $\mu$s long plain MD simulation of BPTI performed on the Anton supercomputer led to identifying four distinct meta-stable states in the structural proximity of the crystallographic native structure\cite{Anton2010}. Key structural differences between these 4 states include (i) the presence of a left-handed Cys14-Cys38 disulfide bridge compared to the right-handed one in the crystal structure, (ii) the rotation of aromatic side chains, in specific TYR10, and (iii) the unfolding of the small $\alpha$-helix located at the N-terminus. 
Representative configurations from each of these metastable states are reported in panel A of Fig. \ref{SI_states_anton_fig}. The color code is consistent with the one adopted in the original paper \cite{Anton2010}.

In order to examine whether our iMapD exploration led to visiting the same states, we first gathered all the configurations generated by iMapD {which satisfied $ Q>0.9$ and RMSD$<2\AA$ from the structures provided by Anton --representative of each state.}
iMapD configurations that satisfy the proximity criterion for more than one 
state were assigned to that with a smallest RMSD distance. These sets of iMapD configurations are reported in panel B of Fig. \ref{SI_states_anton_fig}. According to this criterion we could not find any iMapD configuration near state-3 (purple dot in Fig. \ref{SI_states_anton_fig}A). We stress that this state is by far the shortest-lived among those detected in plain MD simulations, which spend more than 80\% of the simulation time in state-0 and state-1.

\begin{figure}[!t]\centering
 \includegraphics[width=1\textwidth]{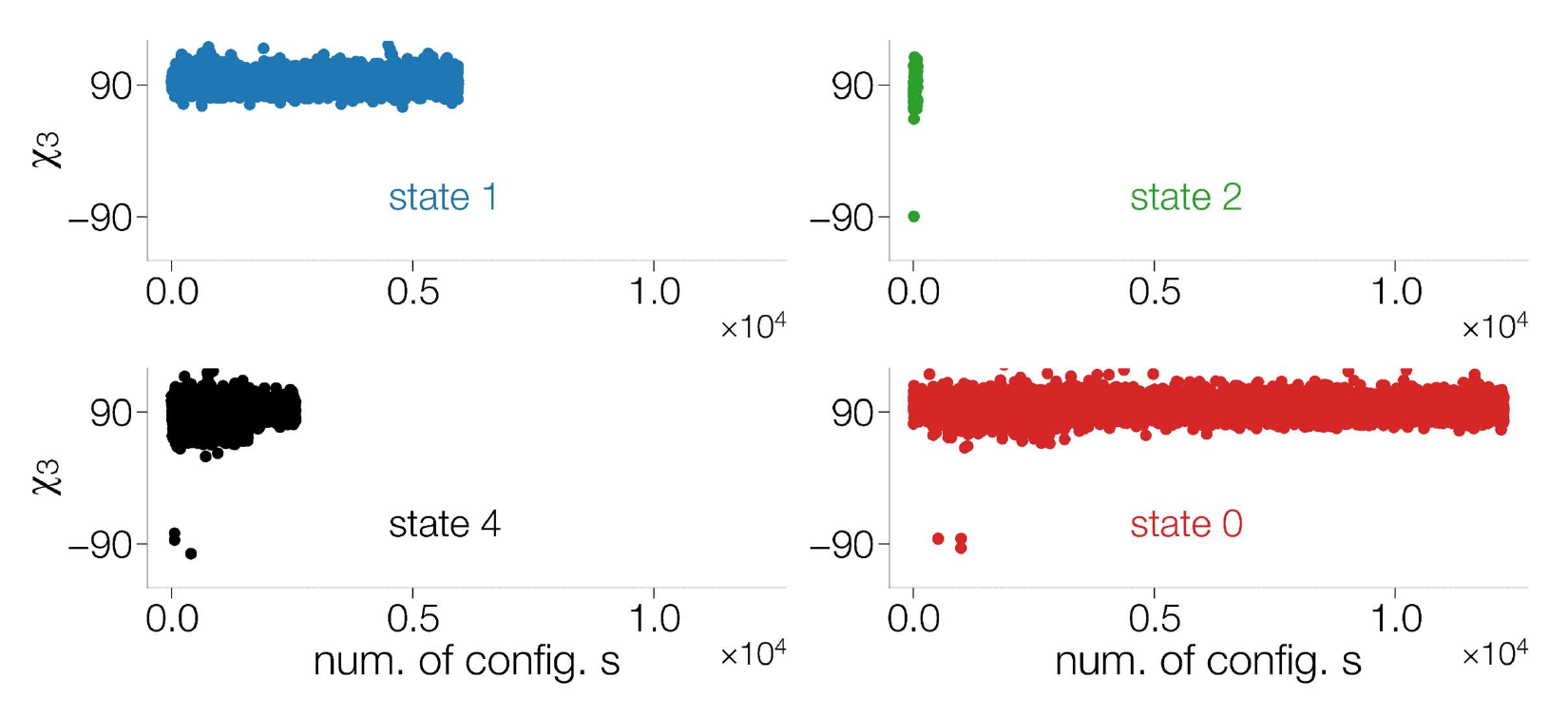}
 \caption{The Cys14-Cys38 bridge dihedral angles ($\chi_3$) for the configurations of iMapD data that correspond to each state of Anton. Consistent with \cite{Anton2010}, we verify that all the states except state-1(crystallographic) contain configurations with left-handed bridges.}\label{SI_chi3_dihedral}
\end{figure}

Since iMapD brakes microscopical reversibility (see discussion in the main text) we could not rely on the dynamical analysis reported in \cite{Anton2010} to assess the meta-stability of the configurations explored by iMapD. We, therefore, relied on an analysis of structural properties to check that the {iMapD configurations associated to each state}
are actually consistent with those found in the metastable states discovered by Anton. In particular, the left-handed chirality of Cys14-Cys38 is one of the main differences between the two largest populated states, state-0 (depicted as the red dot in panel A of Fig. \ref{SI_states_anton_fig}), occupied by $56\%$ of Anton trajectory) and state-1 (the crystallographic state, blue dot in panel A of Fig. \ref{SI_states_anton_fig}, occupied $27\%$ of the time). This property which can be characterized by the dihedral angle $\chi_3\approx-90$, was evaluated for configurations in all the 4 groups of iMapD configurations, as reported in Fig. \ref{SI_chi3_dihedral}. The left-handed disulfide bond in the configurations structurally very close to state-0 provides evidence that iMapD visited this state. Similarly, the characteristic structural properties of state-3 and state-4 observed in \cite{Anton2010} were found in the configurations generated by iMapD, as shown in Fig. \ref{SI_chi3_dihedral}.

\begin{figure}\centering
 \includegraphics[width=1\textwidth]{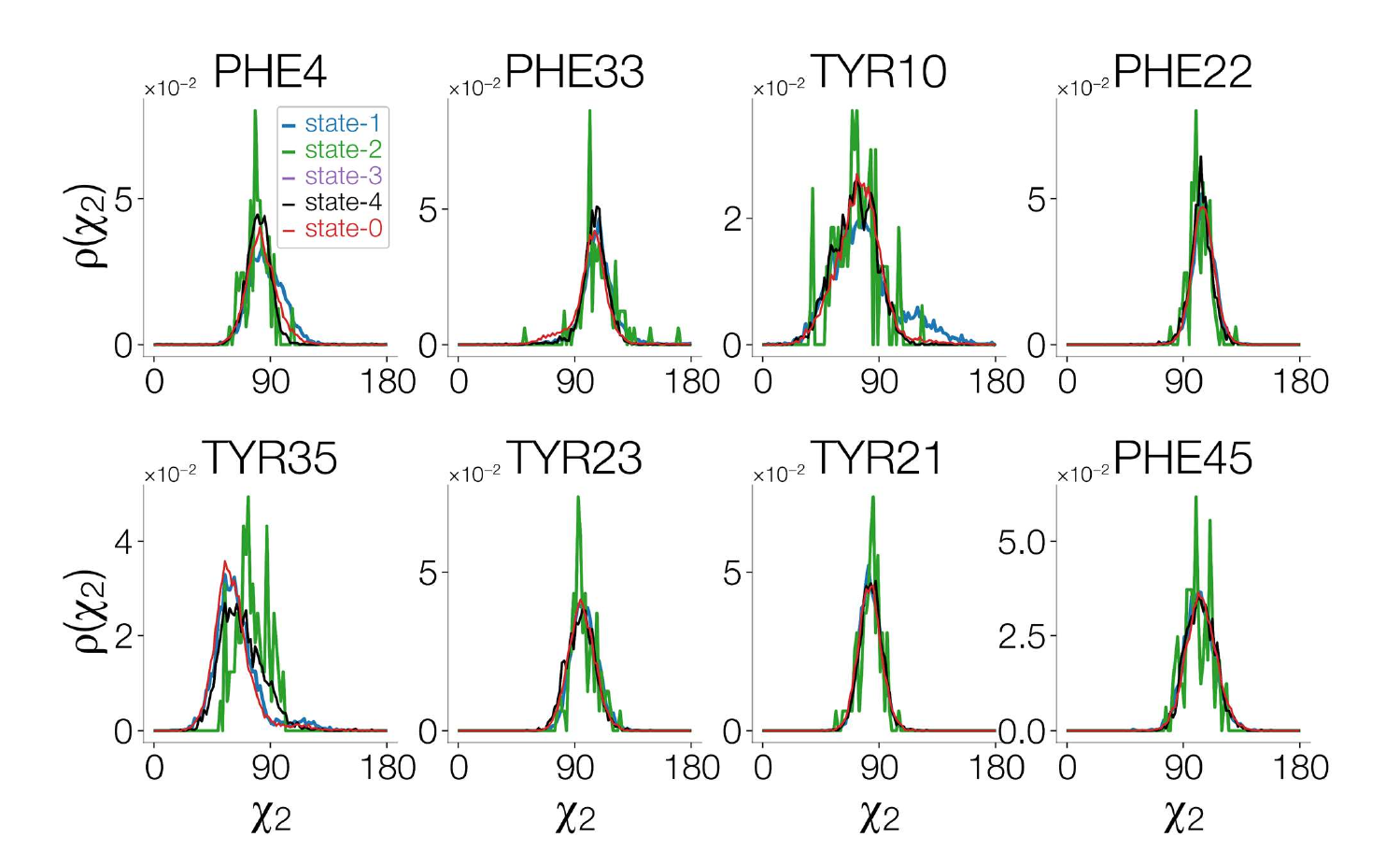}
 \caption{The distribution of $\chi_2$ dihedral angles in certain aromatic side chains of BPTI. This figure corresponds to the same analysis done in \cite{Anton2010}.}\label{SI_chi2_dihedral}
\end{figure}

Another structural analysis, similar to the one reported in \cite{Anton2010}, involved the values of $\chi_2$ dihedral angles in the aromatic side chains. We observed the same angles in the iMapD sets of configurations, as reported in Fig. \ref{SI_chi2_dihedral}.


\section{{Comparing the exploration based on iMapD with one based on high temperature MD}}

\begin{figure}[t!]
\centering
\includegraphics[width=1\textwidth]{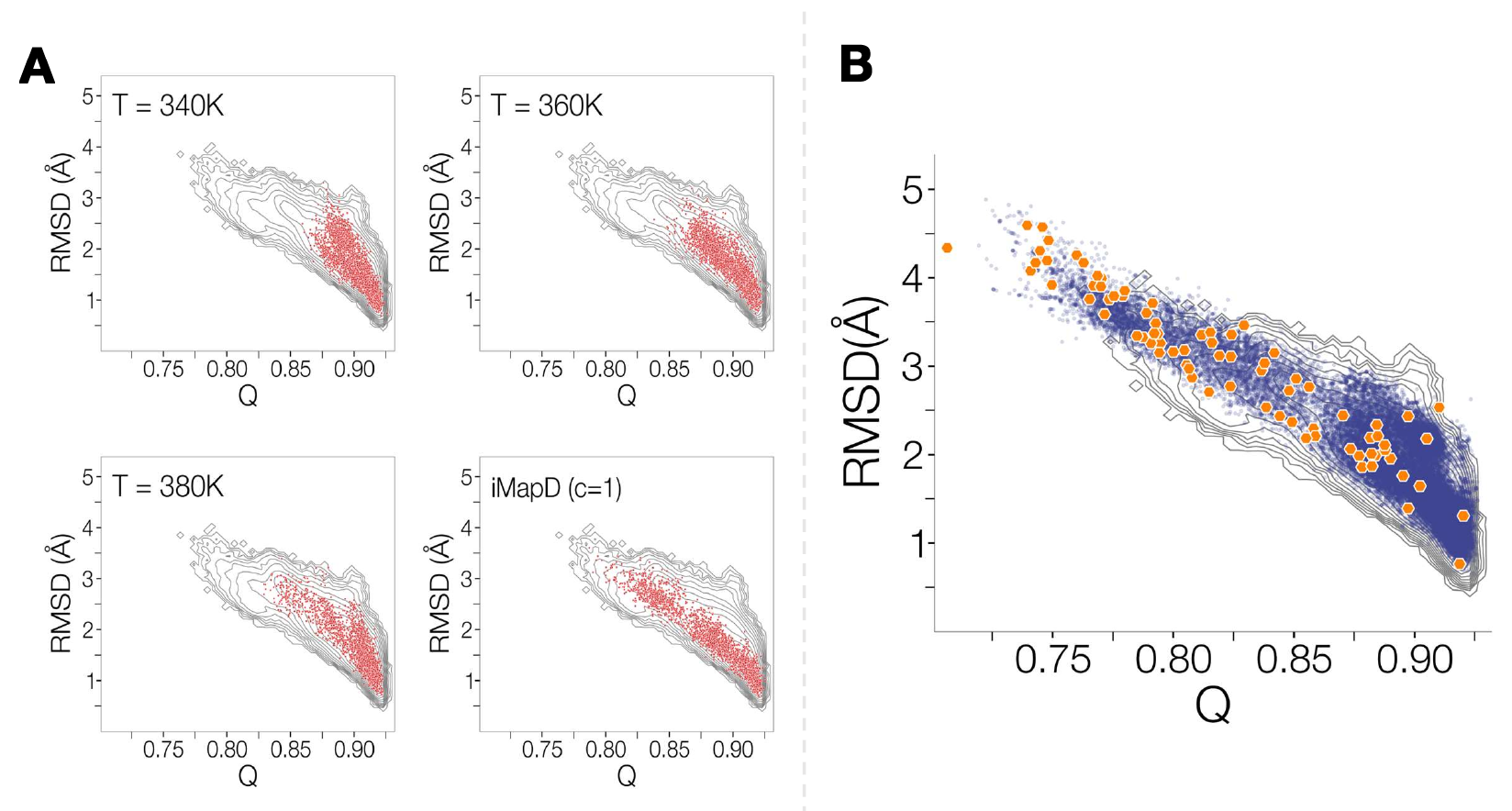}
\caption{Panel A: configurations generated by 300 ns of plain MD at three different high temperatures and 300 ns of cumulative iMapD exploration at ($c=1$), projected onto the Q-RMSD plane. 
 Panel B: The dark blue dots are all the configurations generated during the iMapD exploration performed with three different values of $c$ (0.5, 0.75, and 1.0). The orange dots are the $\nu= 80$ representative centroids obtained after clustering. In all panels, the contour lines in the background highlight the structure of the free-energy landscape evaluated from a frequency histogram of the ms-long plain MD simulation at 300 K, performed with the Anton supercomputer.}\label{fig_iMapD_vs_highTMD}. 
\end{figure}

In this section, we compare the exploration of the intrinsic manifold of protein BPTI performed by iMapD with an exploration based on high-temperature plain MD. To this end, in Fig.\ref{fig_iMapD_vs_highTMD}, we report the results of 300 ns of iMapD (corresponding to the specific choice $c\,=\,1$) with three equally long plain MD simulations, performed at $T\,=\,340,\, 360,$ and $380$ K, respectively. 
All these simulations correctly capture the global structure of the free-energy landscape Q-RMSD plane in the near-native region, i.e. for Q$\,>\,0.9$ and RMSD$\,<\,1.5$\AA. However, the configurations generated by high-temperature MD simulations are confined in a region with Q$\,\gtrsim\, 0.85$, while the iMapD data reach configurations with lower fraction of native contacts and larger RMSD to the native state. Furthermore, for Q$\,<\,0.9$ region, the high-temperature MD configurations drift towards the region with RMSD$\,>\,2\AA$ (except for a branch of data at $T\,=\,380$K), which is scarcely visited at room temperature. 
Conversely, the configurations generated by iMapD (which are based on short MD simulations at $T\,=\,300$K) more faithfully profile the low free-energy regions, all the way to Q$\,<\,0.8$ and RMSD$\,>\,2.5 \AA$. Panel B shows the cumulative results of iMapD exploration based on three different choices of the $c$ parameter, which agrees with the overall structure of the free energy landscape obtained by Anton. The orange points are the 80 representative configurations emerging after clustering and used in constructing the gTPS network.

\section{Additional paths obtained on the DWAVE quantum computer}
\begin{figure}[t!]
\centering
\includegraphics[width=1\textwidth]{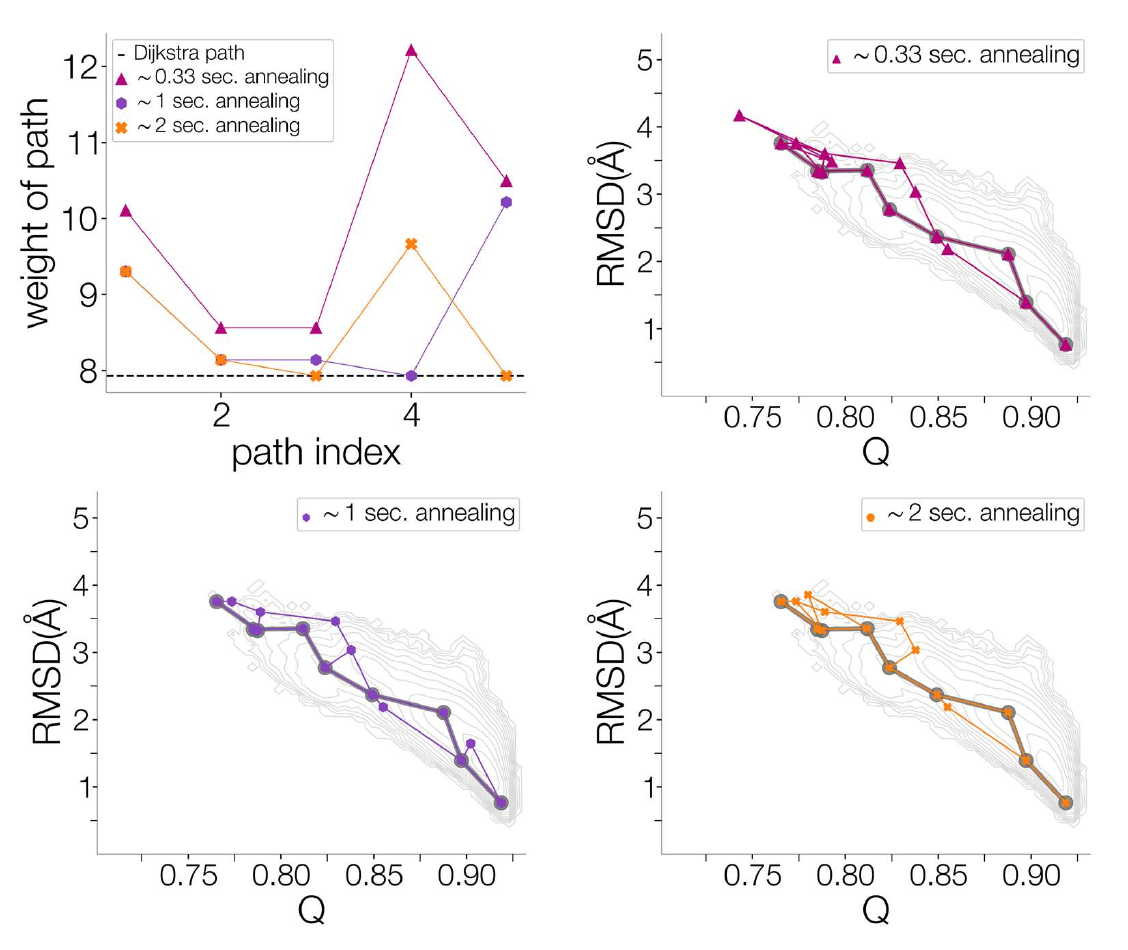}
\caption{Additional transition paths computed using the DWAVE quantum computer and compared with the most probable path obtained with the Dijkstra algorithm. 
Panel A: Evolution of the statistical weight of the transition path (i.e. so-called Hamilton-Jacobi action) along a Markov chain in which the quantum computer generates new paths. Different curves correspond to different sweeping times in the quantum annealing algorithm (see Method section in the main text). }\label{morepaths}. 
\end{figure}

In Fig.\ref{morepaths}, we compare the statistical weight and the structure of stochastic trajectories generated on the DWAVE quantum annealer, using different sweeping times. 

\bibliography{arxiv_preprint}

\end{document}